\documentclass[superscriptaddress,notitlepage,nofootinbib]{revtex4-1}
\usepackage{natbib}
\usepackage{physics}            
\usepackage{amsmath, amsfonts}
\usepackage{siunitx}
\usepackage{graphicx}
\usepackage[english]{babel}
\usepackage[utf8]{inputenc}
\usepackage[autostyle]{csquotes}
\MakeOuterQuote{"}
\usepackage[shortlabels]{enumitem}
\usepackage{dsfont}
\usepackage{bm}
\usepackage{xcolor}
\usepackage{mathtools}
\usepackage[hidelinks]{hyperref}	
\usepackage{framed}
\usepackage[capitalize]{cleveref}
\usepackage{pgf}

\usepackage[normalem]{ulem}
\newcommand\blueout{\bgroup\markoverwith
{\textcolor{blue}{\rule[.5ex]{2pt}{0.4pt}}}\ULon}

\addto{\captionsenglish}{}
\crefname{appsec}{Supplementary Note}{Supplementary Notes}
\Crefname{appsec}{Supplementary Note}{Supplementary Notes}
\crefname{appendix}{Supplementary Note}{Supplementary Notes}
\Crefname{appendix}{Supplementary Note}{Supplementary Notes}

\newcommand{\Scut}{\ensuremath{S_{\textrm{cut}}}}
\newcommand{\Nph}{\ensuremath{N_{\textrm{ph}}}}
\newcommand{\eph}{\ensuremath{e_{\textrm{ph}}}}

\begin{document}
	
	\title{Tight finite-key security for twin-field quantum key distribution}
	\author{Guillermo \surname{Currás Lorenzo}}
	\email{g.j.curraslorenzo@leeds.ac.uk}
	\affiliation{School of Electronic and Electrical Engineering, University of Leeds, Leeds, UK}
	\author{Álvaro \surname{Navarrete}}
	\affiliation{EI Telecomunicación, Dept. of Signal Theory and Communications, University of Vigo, E-36310 Vigo, Spain}
	\author{Koji \surname{Azuma}}
	\affiliation{NTT Basic Research Laboratories, NTT Corporation,
		3-1 Morinosato-Wakamiya, Atsugi, Kanagawa 243-0198, Japan}
	\affiliation{NTT Research Center for Theoretical Quantum Physics, NTT Corporation,
		3-1 Morinosato-Wakamiya, Atsugi, Kanagawa 243-0198, Japan}
	\author{Go \surname{Kato}}
	\affiliation{NTT Research Center for Theoretical Quantum Physics, NTT Corporation,
		3-1 Morinosato-Wakamiya, Atsugi, Kanagawa 243-0198, Japan}
	\affiliation{NTT Communication Science Laboratories, NTT Corporation, 3-1 Morinosato-Wakamiya, Atsugi, Kanagawa 243-0198, Japan}
	\author{Marcos  \surname{Curty}}
	\affiliation{EI Telecomunicación, Dept. of Signal Theory and Communications, University of Vigo, E-36310 Vigo, Spain}
	\author{Mohsen \surname{Razavi}}
	\affiliation{School of Electronic and Electrical Engineering, University of Leeds, Leeds, UK}
	
	\begin{abstract}
Quantum key distribution (QKD) offers a reliable solution to communication problems that require long-term data security. For its widespread use, however, the rate and reach of QKD systems must be improved. Twin-field (TF) QKD is a step forward toward this direction, with early demonstrations suggesting it can beat the current rate-versus-distance records. A recently introduced variant of TF-QKD is particularly suited for experimental implementation, and has been shown to offer a higher key rate than other variants in the asymptotic regime where users exchange an infinite number of signals. Here, we extend the security of this protocol to the finite-key regime, showing that it can overcome the fundamental bounds on point-to-point QKD with around $10^{10}$ transmitted signals. Within distance regimes of interest, our analysis offers higher key rates than those of alternative variants. Moreover, some of the techniques we develop are applicable to the finite-key analysis of other QKD protocols.
\end{abstract}
	
	\maketitle
	
	Quantum key distribution (QKD) enables two remote parties, Alice and Bob, to generate a shared secret key in the presence of an eavesdropper, Eve, who may have unbounded computational power at her disposal \cite{scarani2009security,lo2014secure,pirandola2019advances}. While, ideally, the two parties can be at any distance, in practice, due to the loss and noise in the channel, point-to-point QKD is limited to a certain maximum distance at which secret key bits can securely be exchanged. In fact, the longest distance achieved to date in a terrestrial QKD experiment is about $\SI{400}{\km}$ \cite{yin2016measurement,QKD421km}. The main limitation is the exponential decrease of the transmittance, $\eta$, with the channel length in optical fibres. Even with a high repetition rate of $\SI{10}{\giga \hertz}$, it would take an average of about two minutes to send a single photon over a distance of $\SI{600}{\km}$ of standard optical fibres, and about 300 years to send it over $\SI{1000}{\km}$ \cite{sangouard2011quantum}. Indeed, fundamental bounds \cite{pirandola2009direct,takeoka2014fundamental,pirandola2017fundamental,wilde2017converse,pirandola2018theory} on the private capacity of {\em repeaterless} point-to-point QKD protocols  show that their secret-key rate scales at best approximately linearly with $\eta$. A protocol that aims to overcome this linear scaling must then include at least one middle node. Interestingly, this is not a sufficient condition. A well-known counterexample is the so-called measurement-device independent QKD (MDI-QKD) \cite{lo2012measurement}, which uses the middle node for an {\em untrusted} Bell-state measurement operation. There are, however, extensions of MDI-QKD that can improve its rate scaling from $\eta$ to $\sqrt{\eta}$ by either using quantum memories \cite{panayi2014memory, abruzzo2014measurement} or quantum non-demolition measurements \cite{azuma2015}. Such setups can, in fact, be considered to be the simplest examples of quantum repeaters \cite{duan2001long,sangouard2011quantum}, which are the ultimate solution to trust-free long-distance quantum communications \cite{piparo2014long}. However, even these simple versions may need more time to be efficiently implemented in practice \cite{Lukin_MAQKD,trenyi2019beating}.
	
	Remarkably, the recently proposed twin-field QKD (TF-QKD) \cite{lucamarini2018overcoming} can also overcome this linear scaling while using a relatively simple setup. TF-QKD is related to MDI-QKD, and it inherits its immunity to detector side-channels. However, it relies on single-photon, rather than two-photon, interference for its entanglement swapping operation. The secret-key rate of this protocol was first conjectured \cite{lucamarini2018overcoming} and then proven \cite{tamaki2018information, ma2018phase} to scale with $\sqrt{\eta}$ too, making this approach a strong candidate to beat the current QKD records \cite{minder2019experimental, zhong2019proof, liu2019experimental, wang2019beating} with today's technology. The main experimental challenge is that single-photon interference needs very precise phase stability, which makes it more demanding than two-photon interference. Also, some of its current security proofs \cite{tamaki2018information, ma2018phase} need Alice and Bob to randomly choose a global phase, and then post-select only those rounds in which their choices match, which causes a drop in the secret key rate. Since the original proposal, several variants of TF-QKD have been developed \cite{lin2018simple,curty2019simple,cui2019twin,wang2018twin}, sharing the single-photon interference idea and its consequent $\sqrt{\eta}$ scaling, but differing in their experimental setups and security proofs. Moreover, some of these variants have been shown to be robust against phase misalignment \cite{curty2019simple,cui2019twin,wang2018twin}, which simplifies their experimental implementation. 
	
	In this paper, we focus on the TF-QKD variant introduced in \cite{curty2019simple}, which has two key features: (i) it does not need phase post-selection, which results in a higher secret-key rate; and (ii) it is a convenient option for experimental implementation. Indeed, most of the current TF-QKD experiments use this variant \cite{minder2019experimental,zhong2019proof,wang2019beating}. One of its defining characteristics is its unconventional security proof; specifically, its estimation of the phase-error rate of the protocol, a parameter needed to bound the amount of key information that may have leaked to an eavesdropper. In many QKD protocols, the phase-error rate of the single-photon emissions in one basis can be directly estimated by bounding the bit-error rate of the single-photon emissions in the other basis. In the above TF-QKD variant, however, the encoding bases are not mutually unbiased. To estimate the phase-error rate, the authors in \cite{curty2019simple} use the complementarity \cite{koashi2009simple} between the "phase" and the "photon-number" of a bosonic mode. In this case, the security of a bit encoded in the relative phase of two coherent pulses can be related to the detection statistics of photon-number states. More specifically, in the {\em asymptotic} regime, the phase-error rate can be bounded by a non-linear function of infinitely many yield probabilities for even photon-number states \cite{curty2019simple}, which can be estimated via the decoy-state method \cite{hwang2003quantum,lo2005decoy,wang2005beating}.
	
	
	
	While, in the asymptotic regime, the protocol in \cite{curty2019simple} offers a higher key rate than its counterparts, it is not obvious if this advantage will still hold in a practical setting where only a finite number of pulses is sent. In the finite-key regime, one should account for possible statistical fluctuations between the true phase-error rate and the measurement data used to estimate it. There are, however, two challenges in doing so.
	The first challenge is that the phase-error rate of the protocol is related to the measurement statistics of infinitely many combinations of photon-number states; in practice, one can only obtain bounds for a finite number of them, and dealing with the unbounded components is not as straightforward as in the asymptotic regime. The second challenge is that, unlike in many other QKD protocols, the encoding bases are not mutually unbiased. This opens the possibility that, under a coherent attack by Eve, the detection statistics of a particular round may depend on the basis choices made in previous rounds. Accounting for these correlations makes the analysis quite cumbersome.

	In this work, we provide a rigorous security proof for the protocol in \cite{curty2019simple} that accounts for these two issues in the finite-key setting. Our security proof provides a tight bound on the key rate against general coherent attacks. To overcome the two main challenges mentioned above, we borrow ideas from the finite-key analysis of MDI-QKD \cite{curty2014finite} and the loss-tolerant protocol \cite{tamaki2014losstolerant,mizutani2015finite}, as well as introduce new methods of our own.  To obtain a tighter result, we employ a novel technique to bound the deviation between a sum of correlated random variables and its expected value \cite{kato2020bound}, which can be much tighter than the widely employed Azuma's inequality \cite{azuma1967weighted} when the success probability is low. 
	Importantly, our numerical simulations show that the protocol can overcome the repeaterless bounds \cite{takeoka2014fundamental,pirandola2017fundamental,wilde2017converse} for a block size of only $10^{10}$ transmitted signals {in nominal working conditions.
	
    During the preparation of this manuscript, an alternative finite-key security analysis for an identical protocol setup has been reported in \cite{maeda2019repeaterless}, using an interesting, but different, approach. We would like to highlight that, under identical channel conditions, our analysis results in a higher secret-key rate and imposes fewer conditions on the setup parameters than that of Ref.~\cite{maeda2019repeaterless}. In the Discussion section, we compare both approaches.} We also compare our results with those of the sending-or-not-sending TF-QKD protocol introduced in \cite{wang2018twin}, whose security has recently been extended to the finite-key regime \cite{jiang2019unconditional}. We find that for reasonably large block sizes, the asymptotic key rate advantage of the scheme in \cite{curty2019simple} is maintained in the finite-key regime, for most practical ranges of distance.

	\section*{Results}
	
	\subsection*{Protocol description}
	
\begin{figure}[h]
		\centering
		\includegraphics[width=0.65\textwidth]{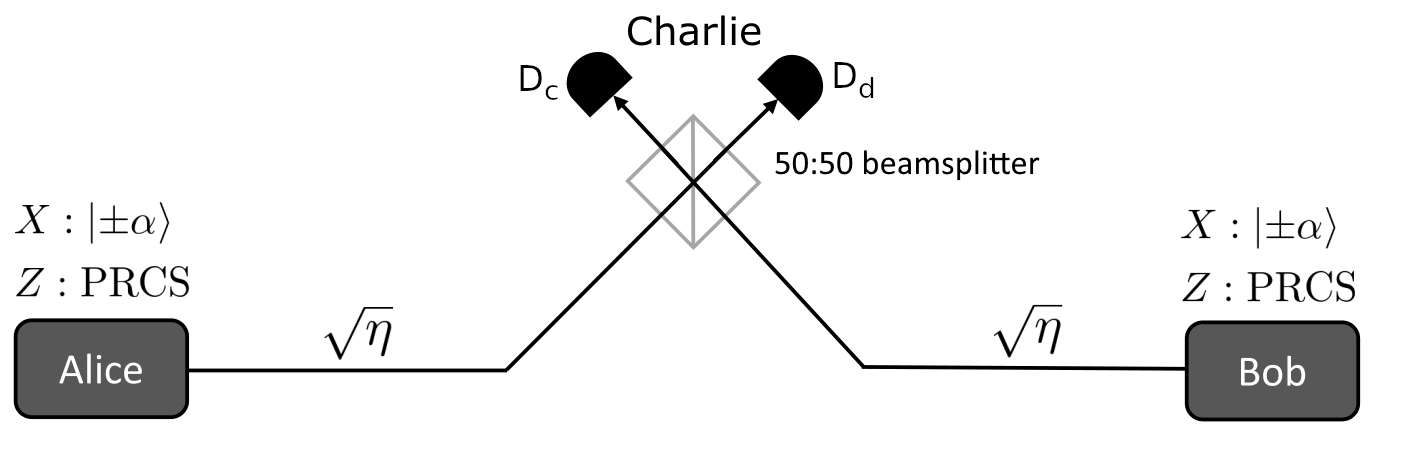}	
		\caption{Setup for the TF-QKD protocol \cite{curty2019simple} described in Box 1. Alice and Bob generate their raw key from the rounds in which they both select the $X$ basis and Charlie declares that a single detector has clicked. The key bit is encoded in the phase of their coherent state. When the users select the same (a different) bit, the constructive (destructive) at Charlie's 50:50 beamsplitter interference should cause a click in detector $D_c$ ($D_d$). The $Z$-basis PRCSs are only used to estimate the phase-error rate of the $X$-basis emissions.  } \label{fig:schematics}
	\end{figure}

	The setup of the TF-QKD protocol in \cite{curty2019simple} is illustrated in \cref{fig:schematics} and its step-by-step description is given in Box~1. Alice and Bob generate quantum signals and send them to a middle node, Charlie, who would ideally couple them at a balanced 50:50 beamsplitter and perform a photodetection measurement. For simplicity, we assume the symmetric scenario in which the Alice-Charlie and Bob-Charlie quantum channels are identical. We note, however, that our analysis can be straightforwardly extended to the asymmetric scenario recently considered in Refs.~\cite{FedAlv,wang2019simple}.  The emitted quantum signals belong to two bases, selected at random. In the $X$ basis, Alice and Bob send phase-locked coherent states $\ket{\pm \alpha}$ with a random phase of either 0 or $\pi$ with respect to a pre-agreed reference. In the $Z$ basis, Alice and Bob generate phase-randomised coherent states (PRCSs), which are diagonal in the Fock basis. The $X$-basis states are used to generate the key, while the $Z$-basis data is used to estimate the detection statistics of Fock states, in combination with the decoy-state method. This is a crucial step in estimating the phase-error rate of the key, thus bounding the information that could have been leaked to a potential eavesdropper. 
	
	\noindent\rule{\textwidth}{1pt}
	\small
	\textbf{Box 1: Simple TF-QKD protocol}
	\begin{enumerate}[label*=(\arabic*)]
		\item \textit{Preparation} 
		
		Alice (Bob) chooses the key-generation basis $X$  with probability $p_X$ or the parameter-estimation basis  $Z$ with probability $p_Z = 1- p_X$, and
		
		\begin{enumerate}[label=(\arabic{enumi}.\arabic*), ref=\arabic{enumi}.\arabic*] 
			\item \label{} If she (he) chooses the $X$ basis, she (he) generates a random bit $b_A$ ($b_B$), prepares an optical pulse in the coherent state $\ket{(-1)^{b_A} \alpha}$ ($\ket{(-1)^{b_B} \alpha}$), and sends it to Charlie.
			
			\item If she (he) chooses the $Z$ basis, she (he) sends an optical pulse in a PRCS of intensity $\mu$, selected from the set $\underline{\mathbf{\mu}}=\{\mu_0,\mu_1,\ldots, \mu_{d-1}\}$  with probability $p_{\mu}$, where $d$ is the number of decoy intensities used.
		\end{enumerate}
		
		They repeat step (1) for $N$ rounds.
		
		\item \textit{Detection} 
		
		An honest Charlie measures each round separately by interfering Alice and Bob's signals at a 50:50 beamsplitter, followed by threshold detectors $D_c$ and $D_d$ placed at the output ports corresponding to constructive and destructive interference, respectively. After the measurement, Charlie reports the pair ($k_c$, $k_d$), where $k_c = 1$ ($k_d = 1$) if detector $D_c$ ($D_d$) clicks and $k_c = 0$ ($k_d = 0$) otherwise. If he is dishonest, Charlie can measure all rounds coherently using an arbitrary quantum measurement, and report $N$ pairs ($k_c$, $k_d$) depending on the result. A round is considered successful (unsuccessful) if $k_c \neq k_d$ ($k_c = k_d$).
		
		\item \textit{Sifting} 
		
		For all successful rounds, Alice and Bob disclose their basis choices, keeping only those in which they have used the same basis. Let $\mathcal{M}_X$ ($\mathcal{M}_Z$) be the set of successful rounds in which both users employed the $X$ ($Z$) basis, and let $M_X = \abs{\mathcal{M}_X}$ ($M_Z = \abs{\mathcal{M}_Z}$) be the size of this set. Alice and Bob disclose their intensity choices for the rounds in $\mathcal{M}_Z$ and learn the number of rounds $M^{\mu \nu}$ in $\mathcal{M}_Z$ in which they selected intensities $\mu \in \underline{\mathbf{\mu}}$ and $\nu \in \underline{\mathbf{\mu}}$, respectively. Also, they generate their sifted keys from the values of $b_A$ and $b_B$ corresponding to the rounds in $\mathcal{M}_X$. For those rounds in which $k_c = 0$ and $k_d = 1$, Bob flips his sifted key bit.

		\item \textit{Parameter estimation}
		
		Alice and Bob apply the decoy-state method to $M^{\mu \nu}$, for $\mu,\nu \in \underline{\mathbf{\mu}}$, obtaining upper-bounds $M_{nm}^{\text{U}}$ on the number of rounds $M_{nm}$ in $\mathcal{M}_Z$ in which they sent $n$ and $m$ photons, respectively. They do this for all $n,m \geq 0$ such that $n+m$ is even and $n+m \leq \Scut$ for a prefixed parameter $\Scut$. Then, they use this data to obtain an upper bound $N_{\textrm{ph}}^{\text{U}}$ on  the number of phase errors, $N_{\textrm{ph}}$, in their sifted keys, and check if the upper bound $e_{\textrm{ph}}^{\text{U}} = N_{\textrm{ph}}^{\text{U}}/M_X$ is lower than a predetermined threshold value. If so, they continue to the last step; otherwise they abort the protocol.
		
		\item \textit{Postprocessing} 
		
		\begin{enumerate}[label=(\arabic{enumi}.\arabic*), ref=\arabic{enumi}.\arabic*] 
			\item  \textit{Error correction}: Alice sends Bob a pre-fixed amount $\lambda_{\textrm{EC}}$ of syndrome information bits through an authenticated public channel, which Bob uses to correct errors in his sifted key.
			
			\item  \textit{Error verification}: Alice and Bob compute a hash of their error-corrected keys using a random universal hash function, and check whether they are equal. If so, they continue to the next step; otherwise, they abort the protocol.
			
			\item \textit{Privacy amplification:} Alice and Bob extract a secret key pair $(S_A, S_B)$ of length $\abs{S_A} = \abs{S_B} = \ell$ from their error-corrected keys using a random two-universal hash function.
		\end{enumerate}
		
	\end{enumerate}
	\normalsize
	\noindent\rule{\textwidth}{1pt}
	\subsection*{Parameter estimation and Secret-key rate analysis}
	
	The main contribution of this work---see Methods for the details---is a procedure to obtain a tight upper-bound $\Nph^{\text{U}}$ on the total number of phase errors $\Nph$ in the finite-key regime for the protocol described in Box 1. Namely, we find that except for an arbitrarily small failure probability $\varepsilon$, it holds that
	\begin{equation}
	\label{eq:NphU}
	N_{\textrm{ph}} \leq N_{\textrm{ph}}^{\text{U}} := \frac{p_X^2}{p_Z^2} \sum_{j=0}^{1} \Bigg[\sum_{\substack{n,m \in \mathds{N}_j \\ n+m \leq \Scut}}  \sqrt{\frac{p_{nm|X}}{p_{nm|Z}}}\sqrt{M_{nm}^{\text{U}} + \Delta_{nm}} + \sqrt{M_Z + \Delta} \sum_{\substack{n,m \in \mathds{N}_j \\ n+m > \Scut}}  \sqrt{\frac{p_{nm|X}}{p_{nm|Z}}} ~\Bigg]^2  + \Delta,
	\end{equation}
	where $p_{nm|X}$ ($p_{nm|Z}$) is the probability that Alice and Bob's joint $X$ ($Z$) basis pulses contain $n$ and $m$ photons, respectively, given by
	\begin{gather} \label{eq:pnmX}
	p_{nm|X} = \abs{\braket{\alpha}{n}}^2 \abs{\braket{\alpha}{m}}^2, \\ \label{eq:pnmZ}
	p_{nm|Z} = \sum_{\mu,\nu \in \underline{\mu}} p_{\mu} p_{\nu}  p_{n|\mu} p_{m|\nu}, 
	\end{gather}
	with $p_{n|\mu} = \mu^n \exp(-\mu)/n!$ being the Poisson probability that a PRCS pulse of intensity $\mu$ will contain $n$ photons; $\Delta$ and $\Delta_{nm}$ are statistical fluctuation terms defined in step 4 of Box 2; $\mathds{N}_0$ ($\mathds{N}_1$) is the set of non-negative even (odd) integers; and the rest of the parameters have been introduced in Box 1. The phase-error rate is then simply upper-bounded by $\eph^{\text{U}} := \Nph^{\text{U}}/M_X$. Box 2 provides a step-by-step instruction list to apply our results to the measurement data obtained in an experimental setup.
	
	When it comes to finite-key analysis, there is one key difference between the protocol in Box 1 and several other protocols, such as, for example, decoy-state BB84 \cite{lim2014concise}, decoy-state MDI-QKD \cite{curty2014finite}, and sending-or-not-sending TF-QKD \cite{jiang2019unconditional}. In all the latter setups, when there are no state-preparation flaws, the single-photon components of the two encoding bases are mutually unbiased; in other words, they look identical to Eve once averaged by the bit selection probabilities. This implies that such states could have been generated from a maximally entangled bipartite state, where one of its components is measured in one of the two orthogonal bases, and the other half represents an encoded key bit. In fact, the user(s) could even wait until they learn which rounds have been successfully detected to decide their measurement basis, effectively delaying their choice of encoding basis. This possibility allows the application of a random sampling argument: since the choice of the encoding basis is independent of Eve's attack, the bit error rate of the successful $X$-basis emissions provides a random sample of the phase-error rate of the successful $Z$-basis emissions, and vice-versa. Then, one can apply tight statistical results such as the Serfling inequality \cite{serfling1974probability} to bound the phase-error rate in one basis using the measured bit-error rate in the other basis. This approach, however, is not directly applicable to the protocol in Box 1, in which the secret key is extracted from all successfully detected $X$-basis signals, not just from their single-photon components. Moreover, the encoding bases are not mutually unbiased: the $Z$-basis states are diagonal in the Fock basis, while the $X$-basis states are not. This will require a different, perhaps more cumbersome, analysis as we highlight below.
	
	To estimate the $X$-basis phase-error rate from the $Z$-basis measurement data, we construct a virtual protocol (see Box 3) in which the users learn their basis choice by measuring a quantum coin after Charlie/Eve reveals which rounds were successful. Note that, because of the biased basis feature of the protocol, the statistics of the quantum coins associated to the successful rounds could depend on Eve's attack. This means that the users cannot delay their choice of basis, which prevents us from applying the random sampling argument. Still, it turns out that the quantum coin technique now allows us to upper-bound the average number of successful rounds in which the users had selected the $X$ basis and undergone a phase error. This bound is a non-linear function of the average number of successful rounds in which they had selected the $Z$ basis and respectively sent $n$ and $m$ photons, with $n+m$ even. More details can be found in the Methods Section; see \cref{eq:Nph-intermediate}.
	
	The main tool we use to relate each of the above average terms to their actual occurrences, $\Nph$ and $M_{nm}$, is Azuma's inequality \cite{azuma1967weighted}, which is widely used in security analyses of QKD to bound sums of observables over a set of rounds of the protocol (in our case, the set of successful rounds after sifting), when the independence between the observables corresponding to different rounds cannot be guaranteed. When using Azuma's inequality, the deviation term $\Delta$ scales with the square root of the number of terms in the sum. In our case, $\Delta$ scales with $\sqrt{M_s}$, where $M_s$ is the number of successful rounds after sifting. For parameters of comparable magnitude to $M_s$, this provides us with a reasonably tight bound. Whenever the parameter of interest is small, however, the provided bound could instead be loose. This is the case for the crucial term $M_{00}^{\rm U}$ in \cref{eq:NphU}, as vacuum states are unlikely to result in successful detection events, thus the bound obtained with Azuma's inequality can be loose. This is important because, in \cref{eq:NphU}, the coefficient associated to the vacuum term is typically the largest. It is then essential to find a tighter bound for this term. 
	 
	For this, we employ a remarkable recent technique to bound the deviation between a sum of dependent random variables and its expected value \cite{kato2020bound}. This technique provides a much tighter bound than Azuma's inequality when the value of the sum is much lower than the number of terms in the sum. In particular, it provides a tight upper-bound for the vacuum component $M_{00}$. In Methods, we provide a statement of the result and we explain how we apply it to our protocol.

	Having obtained $\eph^{\text{U}}$, we show in \cref{app:security-bounds} that the secret-key length, $\ell$, can be lower bounded by
	\begin{equation}
	\label{eq:keyrate}
	\ell \geq  M_X \left[1-h(e_{\textrm{ph}}^{\text{U}})\right]  - \lambda_{\textrm{EC}} - \log_2 \frac{2}{\epsilon_{\textrm{c}}} - \log_2 \frac{1}{4\epsilon_{\textrm{PA}}^2 },
	\end{equation}
	while guaranteeing that the protocol is $\epsilon_{\textrm{c}}$-correct and $\epsilon_{\textrm{s}}$-secret, with $\epsilon_{\textrm{s}} = 2 \varepsilon + \epsilon_{\textrm{PA}}$;  where $h(x) = -x \log_2 x - (1-x) \log_2 (1-x)$ is the Shannon binary entropy function, $\lambda_{\textrm{EC}}$ is number of bits that are spent in the error-correction procedure, $\epsilon_{\textrm{PA}}$ is the failure probability of the privacy amplification scheme, and $\varepsilon$ is the failure probability associated to the estimation of the phase-error rate. Here, our security analysis follows the universal composable security framework \cite{ben2005universal,renner2005universally}, according to which a protocol is $\epsilon_{\textrm{sec}}$-secure if it is both $\epsilon_{\textrm{c}}$-correct and $\epsilon_{\textrm{s}}$-secret, with  $\epsilon_{\textrm{sec}} \geq \epsilon_{\textrm{c}} + \epsilon_{\textrm{s}}$. The correctness criterion is met when Alice and Bob's secret keys $S_A$ and $S_B$ are identical, and the protocol is $\epsilon_{\textrm{c}}$-correct when $\Pr[S_A \neq S_B] \leq \epsilon_{\textrm{c}}$. The secrecy criterion is met when the classical-quantum state $\rho_{AE}$ describing Alice's secret key and Eve's side information is of the form $\rho_{AE} = U_A \otimes \rho_E$, where $U_A$ is the uniform distribution over all bit strings, and $\rho_E$ is an arbitrary quantum state. The protocol is $\epsilon_{\textrm{s}}$-secret if
	\begin{equation}
	\frac{1}{2} \norm{\rho_{AE} - U_A \otimes \rho_E} \leq \epsilon_{\textrm{s}},
	\end{equation}
	where $\norm{\cdot}$ is the trace norm.

	\noindent\rule{\textwidth}{1pt}
	\small
	\textbf{Box 2: Instructions for experimentalists}
	
	\begin{enumerate}
	    \item Set the security parameters $\epsilon_{\textrm{c}}$ and $\epsilon_{\textrm{PA}}$, as well as the failure probabilities  $\varepsilon_c$ and $\varepsilon_a$ for the inverse multiplicative Chernoff bound and the concentration bounds for sums of dependent random variables, respectively. Calculate the overall failure probability $\varepsilon$ of the parameter estimation process, which depends on the number of times that the previous two inequalities are applied. In general, $\varepsilon = d^2 \varepsilon_c + \left(\lfloor\frac{\Scut}{2}\rfloor+1\right)^2 \varepsilon_a + \varepsilon_a$, where $d$ is the number of decoy intensities employed by each user. For $\Scut = 4$ and three decoy intensities, we have that $\varepsilon = 9 \varepsilon_c + 10 \varepsilon_a$.
	    \item Use prior information about the channel to obtain a prediction $\tilde{M}_{00}^{\rm U}$ on $M_{00}^{\rm U}$, the upper bound on the number of $Z$-basis vacuum events that will be obtained after applying the decoy-state method.
		\item Run steps 1-3 of the protocol in Box 1, obtaining a sifted key of length $M_X$, and $Z$-basis measurement counts $M^{\mu \nu}$ for $\mu,\nu \in \underline{\mathbf{\mu}}$. Let $M_s = M_X + M_Z$ be the number of successful rounds after sifting.
		\item Use the analytical decoy-state method included in the \cref{app:decoy-analysis} to obtain upper bounds $M_{nm}^{\text{U}}$ from $M^{\mu \nu}$. Alternatively, use the numerical estimation method introduced in the Supplementary Notes of \cite{curty2014finite}.
		\item Set $\Delta = \sqrt{ \frac{1}{2} M_s \ln \varepsilon_a^{-1}}$ and $\Delta_{nm} = \Delta$ for all $n,m$ except for $m=n=0$. Substitute $\tilde{\Lambda}_n \to \tilde{M}_{00}^{\rm U}$ in \cref{eq:ab-solution} to find parameters $a$ and $b$. Set
		\begin{equation}
	\Delta_{00} = \left[b+a \left(\frac{2 M_{00}^{\rm U}}{n} - 1 \right) \right] \sqrt{n},
	\end{equation}
		
		\item Use \cref{eq:NphU} to find $\Nph^{\text{U}}$ and set $\eph^{\text{U}} = \Nph^{\text{U}}/M_X$.
		
		\item Use \cref{eq:keyrate} to specify the required amount of privacy amplification and to find the corresponding length of the secret key that can be extracted. The key obtained is $\epsilon_{\textrm{sec}}$-secure, with $\epsilon_{\textrm{sec}} \geq  \epsilon_{\textrm{c}} + \epsilon_{\textrm{s}}$ and $\epsilon_{\textrm{s}} = 2 \varepsilon + \epsilon_{\textrm{PA}}$.
		
	\end{enumerate}
	
	\normalsize
	\noindent\rule{\textwidth}{1pt}
	
	%
	%
	%
	%
	%
	%
	%
	%
	\section*{Discussion}
	
	In this section, we analyse the behaviour of the secret-key rate as a function of the total loss. We simulate the nominal no-Eve scenario with an honest Charlie. In this case, the total Alice-Bob loss includes the loss in the quantum channels as well as the inefficiency of Charlie's detectors. We compare the key rate for the protocol in Box 1, using the finite-key security analysis introduced in the previous section, with that of the sending-or-not-sending TF-QKD protocol \cite{wang2018twin,jiang2019unconditional}, as well as with the finite-key analysis presented in Ref.~\cite{maeda2019repeaterless}. We  also  include  the asymptotic secret key capacity for repeaterless QKD systems over lossy channels, known as the PLOB bound \cite{pirandola2017fundamental}, for comparison. While specific bounds for the finite-key setting have recently been studied \cite{wilde2017converse,laurenza2019tight}, in the practical regimes of interest to this work, they numerically offer a negligible difference to the PLOB bound. The latter has then been used in all relevant graphs for consistency. To simulate the data that would be obtained in all protocols, we use the simple channel model described in \cref{app:channel-model}, which accounts for phase and polarisation misalignments. Also, we assume that both users employ three decoy-state intensities $\mu_0 > \mu_1 > \mu_2$. Since the optimal value $\mu_2 = 0$ is typically difficult to achieve in practice, we set $\mu_2 = 10^{-4}$ and optimise the secret-key rate over the value of $\mu_0$ and $\mu_1$. We also optimise it over the selection probabilities, as well as over $p_X$ and $\alpha$. 
	
		\begin{table}[h]
		\begin{minipage}[c]{0.35\textwidth}
			\centering
			\begin{ruledtabular}
				\begin{tabular}{cccccc}
					$p_d$  & $\delta_{\textrm{ph}}$ & $\delta_{\rm pol}$ & $f$ & $\epsilon_c$ & $\epsilon_s$   \\ \hline 
					$10^{-8}$ & $9.1 \%$ &   $0 $ & 1.16 & $10^{-10}$ & $10^{-10}$
				\end{tabular}
			\end{ruledtabular}
		\end{minipage} 
		\caption{List of parameters used in the numerical simulations. Here, $p_d$ is the dark count probability, per pulse, of the detectors, $\delta_{\textrm{ph}}$ is the phase misalignment of the system, $\delta_{\textrm{pol}}$ is the polarisation misalignment of the system, and $f$ is the error-correction inefficiency. In our numerical simulations, we set  $\varepsilon = \epsilon_{\textrm{PA}} =  \epsilon_{\textrm{s}}/3$.}
		\label{tab:parameters}
	\end{table}
	
	The nominal values for system parameters are summarised in \cref{tab:parameters}. We assume a phase mismatch of 9.1\% between Alice and Bob's signals, corresponding to a QBER of around 2\% for most attenuations, matching the experimental results in \cite{minder2019experimental}. For brevity, we do not consider the effect of polarisation misalignment in our numerical results, but one can use the provided analytical model to study different scenarios of interest. In principle, even if the mechanism used for polarisation stability is not perfect, one can use polarisation filters to ensure that the same polarisation modes are being coupled at the 50:50 beamsplitter, at the cost of introducing additional loss. We assume an error correction leakage of $\lambda_{\textrm{EC}} = f M_X h(e_X)$, where $e_X$ is the bit error rate of the sifted key, and $f$ is the error correction inefficiency.  For the security bounds, we set $\epsilon_{\textrm{c}} = \epsilon_{\textrm{s}} = 10^{-10}$, and for simplicity we set $\varepsilon = \epsilon_{\textrm{PA}} =  \epsilon_{\textrm{s}}/3$.
	
	\begin{figure}[h]
		\centering
    	\resizebox{.8\linewidth}{!}{\input{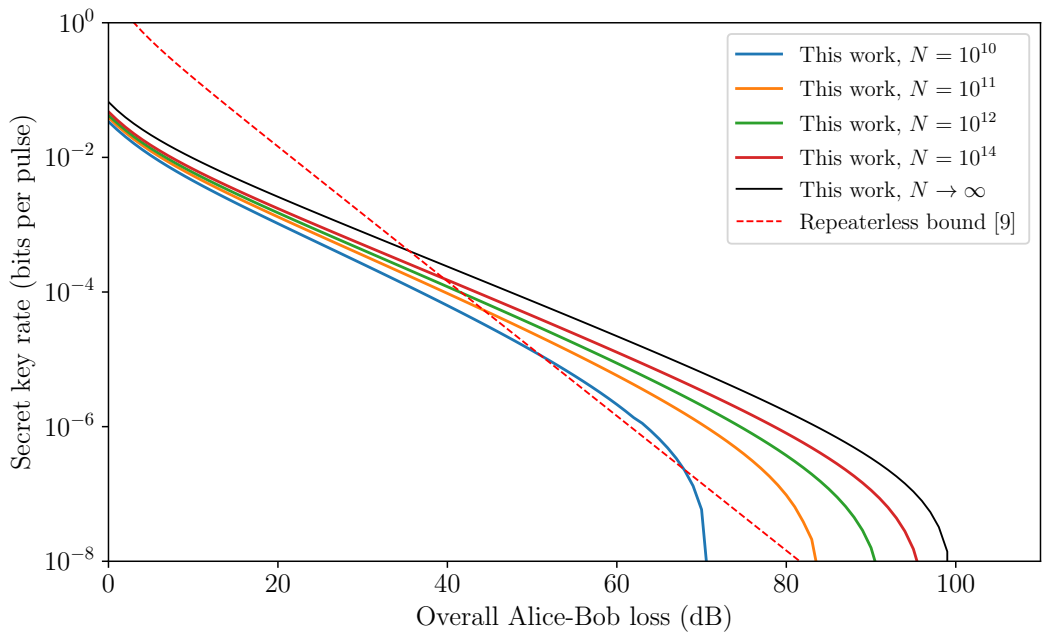}}
		\caption{Secret key rate per pulse for the protocol in Box 1 for different values of the block size, $N$, which represents the total number of rounds in the protocol. The overall Alice-Bob loss includes the loss in the quantum channels and Charlie's detectors. The simulation parameters are listed in \cref{tab:parameters}.}
		\label{fig:graph1}
	\end{figure}
	
	\begin{figure}[h]
		\centering
    	\resizebox{.8\linewidth}{!}{\input{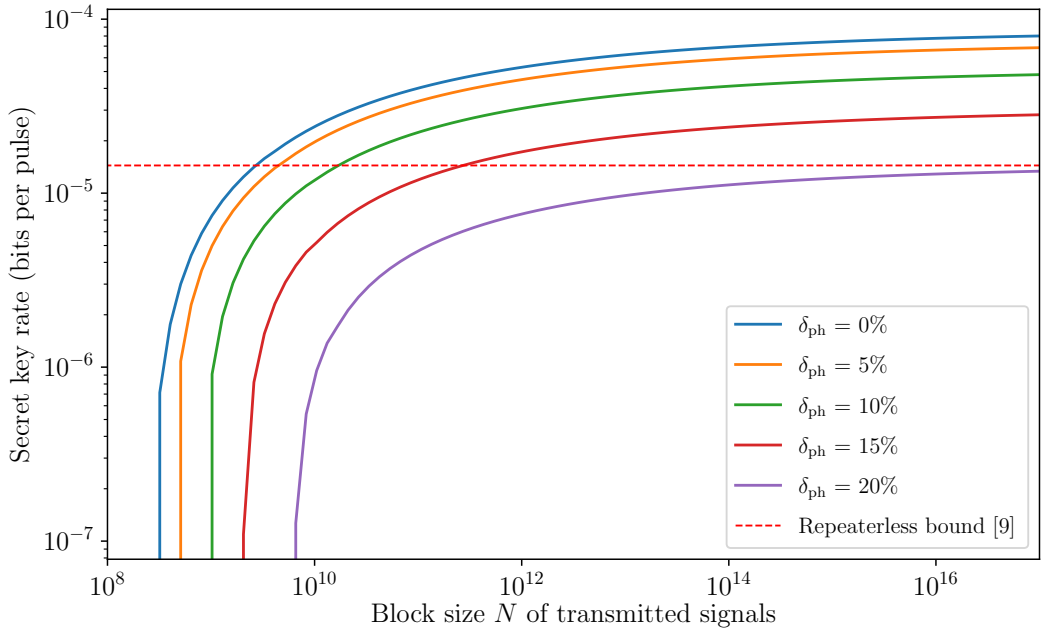}}
		\caption{Secret key rate per pulse for the protocol in Box 1 at a total loss of $\SI{50}{\decibel}$, for different values of phase misalignment, as a function of the block size $N$. All other simulation parameters are listed in \cref{tab:parameters}.}
		\label{fig:graph2}
	\end{figure}
	
	\begin{figure}[h]
		\centering
	    \resizebox{.8\linewidth}{!}{\input{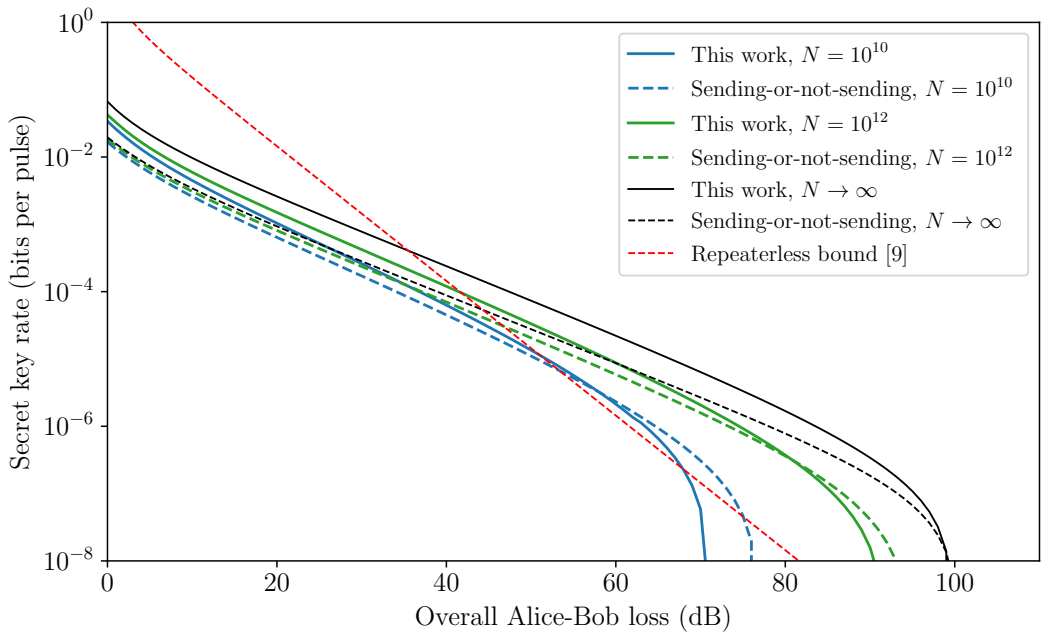}}
		\caption{Comparison between this work and sending-or-not-sending TF-QKD\cite{wang2018twin,jiang2019unconditional}, for different block size values $N$ of transmitted signals. All other simulation parameters are listed in \cref{tab:parameters}.}
		\label{fig:graph3}
	\end{figure}

	\begin{figure}[h]
	\centering
	\resizebox{.8\linewidth}{!}{\input{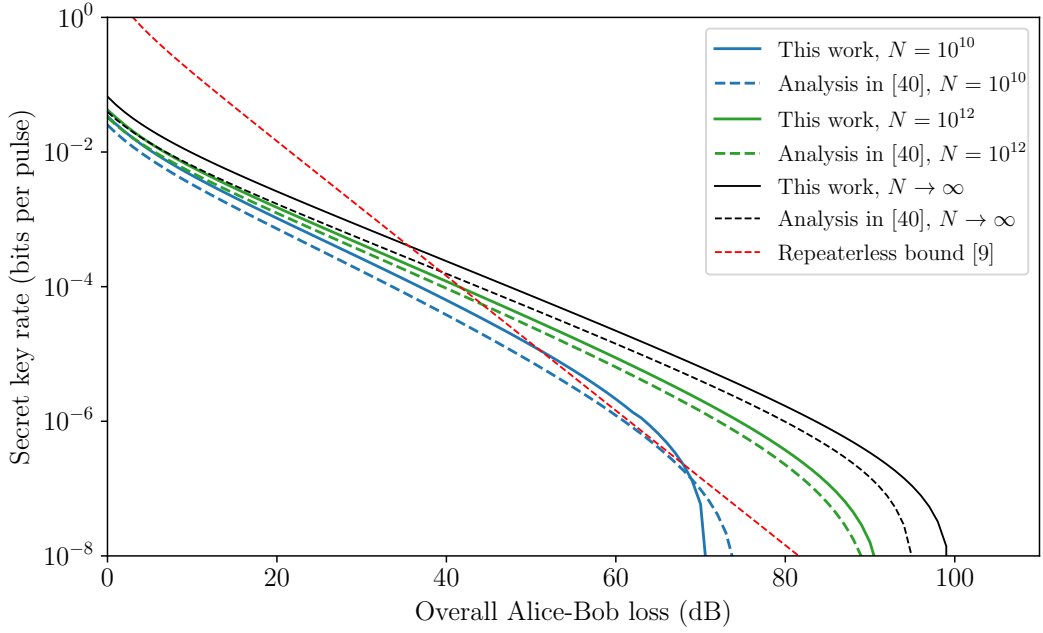}}
	\caption{Comparison between this work and the alternative analysis in \cite{maeda2019repeaterless}, for different block size values $N$ of transmitted signals. The simulation parameters are listed in \cref{tab:parameters}.}
	\label{fig:graph4}
\end{figure}
	
	In \cref{fig:graph1}, we display the secret key rate per pulse for the protocol in Box 1 for different values of the block size, $N$, of transmitted signals. It can be seen that the protocol can outperform the repeaterless bound for a block size of just $10^{10}$ transmitted signals per user, at an approximate total loss of 50 dB. For standard optical fibres, this corresponds to a total distance of 250~km, if we neglect the loss in the photodetectors. At a 1~GHz clock rate, it takes only ten seconds to collect the required data. For a block size of $10^{11}$ transmitted signals, the protocol can already outperform the repeaterless bound for a total loss ranging from $\SI{45}{\decibel}$ to over $ \SI{80}{\decibel}$. By increasing $N$, we approach the asymptotic performance of the protocol. 
	
    The dependence of the secret key rate on the block size $N$ has been shown in \cref{fig:graph2}, at a fixed total loss of $\SI{50}{\decibel}$ and for several values of phase misalignment. In all cases, there is a minimum required block size to obtain a positive key rate. This minimum block size can be even lower than $10^9$ in the ideal case of no phase misalignment, and it goes up to around $10^{10}$ at $\delta_{\rm ph} =20 \%$. There is a sharp increase in the secret key rate once one goes over this minimum required block size after which one slowly approaches the key rate in the asymptotic limit. The latter behaviour is likely due to the use of Azuma's inequality. One can, nevertheless, overcome the repeaterless bound at a reasonable block size in a practical regime where $\delta_{\rm ph} \leq 15 \%$. At higher values of total loss this crossover happens at even larger values of $\delta_{\rm ph}$. 
	
	In \cref{fig:graph3}, we compare the performance of the protocol in Box 1 with that of the sending-or-not-sending TF-QKD protocol presented in \cite{wang2018twin,jiang2019unconditional}. In the asymptotic regime, the former protocol outperforms the latter at all values of total loss. For a block size of $10^{12}$ transmitted signals, this is still the case up to $\SI{80}{\decibel}$ of total loss, after which the key rate is perhaps too low to be of practical relevance. For a block size of $10^{10}$ transmitted signals, however, the curves for the two protocols cross at around $\SI{55}{\decibel}$, where they happen to cross the repeaterless bound as well. In this case, the sending-or-not-sending protocol offers a better performance after this point. This behaviour is due to the different statistical fluctuation analyses applied to the two protocols. As explained in the Result section, the single-photon components in the sending-or-not-sending protocol are mutually unbiased, allowing for a simpler and tighter estimation of the phase-error rate. This is not the case for the TF-QKD protocol in Box 1, for which this estimation involves the application of somewhat looser bounds for several terms in \cref{eq:NphU}. We conclude that for sufficiently large block sizes, the protocol in Box 1 maintains its better performance over the sending-or-not sending variant.
	
	Finally, in \cref{fig:graph4}, we compare our results with those of the alternative analysis in \cite{maeda2019repeaterless}. To compute the secret-key rate of the latter, we use the code provided by the authors, except for the adjustments needed to match it to the channel model described in \cref{app:channel-model}. It can be seen that, at all cases considered and for most practical regimes of interest, the analysis introduced in this paper provides a higher key rate than that of \cite{maeda2019repeaterless}. Moreover, we remark that the security proof presented in \cite{maeda2019repeaterless}, in its current form, is only applicable when the state generated by the weakest decoy intensity $\mu_2$ is a perfect vacuum state of intensity $\mu_2 = 0$. The security analysis presented in this work, however, can be applied to any experimental value of $\mu_2$, and we assume a value of $\mu_2 = 10^{-4}$, which may be easier to achieve in practice. That said, the security proof in  \cite{maeda2019repeaterless} adopts an interesting approach that results in a somehow simpler statistical analysis. In particular, unlike in the analysis presented in this paper, the authors in \cite{maeda2019repeaterless} do not estimate the detection statistics of photon-number states as an intermediate step to bounding the phase-error rate. Instead, they show that the operator corresponding to a phase-error can be bounded by a linear combination of the $Z$-basis decoy states. While this linear bound is asymptotically looser than the non-linear formula in \cref{eq:NphU}, it allows the application of a simpler statistical analysis based on a double use of Bernoulli sampling. Given that the finite-key analysis of a protocol could be part of the software package of a product, we believe that the additional key rate achievable by our analysis justifies its slightly more complex approach.
	
	In conclusion, we have proven the security of the protocol in Box 1 in the finite-key regime scenario against coherent attacks. Our results show that, under nominal working conditions experimentally achievable by today's technology, this scheme could outperform the repeaterless secret-key rate bound in a key exchange run of only ten seconds, assuming a $\SI{1}{\giga \hertz}$ clock rate. It would also outperform other TF-QKD variants, as well as alternative security proofs in practical regimes of interest. 
	
	\section*{Methods}
	In this section, we introduce the procedure that we use to bound the phase-error rate of the protocol in Box 1, referring to the Supplementary Notes when appropriate. For notation clarity, we assume the symmetric scenario in which Alice and Bob employ the same $X$-basis amplitude $\alpha$ and the same $Z$-basis intensities $\mu \in \underline{\bf \mu}$. However, the analysis can be applied as well to the asymmetric scenario \cite{FedAlv,wang2019simple} by appropriately redefining the parameters $p_{nm|X}$ and $p_{nm|Z}$.
	
	\subsection*{Virtual protocol}
	
	To bound the phase-error rate, we construct a virtual protocol in which Alice and Bob measure an observable that is conjugate to that used to generate the key. By the complementarity argument \cite{koashi2009simple}, the bit-error rate of this virtual protocol is identical to the phase-error rate of the actual protocol, provided that the two protocols are equivalent. The equivalence condition is satisfied if the two protocols send the same quantum and classical information, thus Eve cannot tell which of the two is being performed. More concretely, our virtual protocol replaces Alice's $X$-basis emissions by the preparation of the state
	\begin{equation}
	\label{eq:psix}
	\ket{\psi_x}_{Aa} = \frac{1}{\sqrt{2}} (\ket{+}_{A} \ket{\alpha}_{a} + \ket{-}_{A} \ket{-\alpha}_{a}),
	\end{equation}
	where $A$ is an ancilla system at Alice's lab, $a$ is the photonic system sent to Eve, and $\ket{\pm} = \frac{1}{\sqrt{2}} (\ket{0} \pm \ket{1})$; while Bob's $X$ basis emissions are replaced by a similarly defined $\ket{\psi_x}_{Bb}$. After Eve's attack, Alice and Bob measure systems $A$ and $B$ in the $Z$ basis $\{\ket{0},\ket{1}\}$, which is conjugate to the $X$ basis $\{\ket{+},\ket{-}\}$ that they would use to generate the key. It is useful to write the state in \cref{eq:psix} as
	\begin{equation}
	\ket{\psi_x}_{Aa}  = \ket{0}_{A} \ket{C_0}_{a}+ \ket{1}_{A} \ket{C_1}_{a},
	\end{equation}
	where $\ket{C_0}$ and $\ket{C_1}$ are the (unnormalised) cat states 
	\begin{equation} \label{eq:complementary}
	\ket{C_0} =  \frac{1}{2} (\ket{\alpha} + \ket{-\alpha}),  \qquad \quad \ket{C_1}= \frac{1}{2}(\ket{\alpha} - \ket{-\alpha}).
	\end{equation}
	
	Alice's $Z$-basis emissions are diagonal in the Fock basis, and the virtual protocol replaces them by their purification
	\begin{equation}
	\ket{\psi_z}_{Aa} =  \sum_{n=0}^{\infty}  \sqrt{p_{n|Z}}  \ket{n}_A \ket{n}_a,
	\end{equation}
	where $p_{n|Z} = \sum_{\mu \in \underline{\bf \mu}} p_{\mu} p_{n|\mu}$ is the probability that Alice's $Z$ basis pulse contains $n$ photons, averaged over the selection of $\mu$. Unlike in the actual protocol, in the virtual protocol Alice and Bob learn the photon number of their signals by measuring systems $A$ and $B$ after Eve's attack.

	Lastly, Alice's emission of $\ket{\psi_x}_{Aa}$ with probability $p_X$ and $\ket{\psi_z}_{Aa}$ with probability $p_Z$ is replaced by the generation of the state
	\begin{equation}
	\label{eq:vir_psi}
	\ket{\psi}_{A_cAa}= \sqrt{p_X} \ket{0}_{A_c} \ket{\psi_x}_{A a} + \sqrt{p_Z} \ket{1}_{A_c} \ket{\psi_z}_{A a},
	\end{equation}
	where $A_c$ is a quantum coin ancilla at Alice's lab; while Bob's is replaced by an equally defined $\ket{\psi}_{B_cBb}$. Alice and Bob measure systems $A_c$ and $B_c$ after Eve's attack, delaying the reveal of their basis choice. The different steps of the virtual protocol are described in Box 3.
	
	\noindent\rule{\textwidth}{1pt}
	\small
	\textbf{Box 3: Virtual protocol}
	\begin{enumerate}[label*=(\arabic*), ref=\arabic*]
		\item \textit{Preparation} 
		
		Alice and Bob prepare $N$ copies of the state $\ket{\phi} = \ket{\psi}_{A_cAa} \otimes \ket{\psi}_{B_cBb}$ and send all systems $a$ and $b$ to Eve over the quantum channel.
		
		\item \textit{Detection}  \label{step:detection}
		
		Eve performs an arbitrary general measurement on all the subsystems $a$ and $b$ of $\ket{\phi}^{\otimes N}$ and publicly announces $N$ bit pairs $(k_c, k_d)$. Without loss of generality, we assume that there is a one-to-one correspondence between her measurement outcome and her set of announcements. A round is considered successful (unsuccessful) if $k_c \neq k_d$ ($k_c = k_d$). Let $\mathcal{M}$ ($\bar{\mathcal{M}}$) represent the set of successful (unsuccessful) rounds.
		
		\item \textit{Virtual sifting}  \label{step:virtualsifting}
		
		For all rounds, Alice and Bob jointly measure the systems $A_c$ and $B_c$, learning whether they used the same or different bases, but not the specific basis they used. Let $\mathcal{M}_s$ ($\mathcal{M}_d$) denote the set of  successful rounds in which they used the same (different) bases.
		
		\item \textit{Ancilla measurement} \label{step:ancilla-measurement}
		
		\begin{enumerate}[label=(\arabic{enumi}.\arabic*), ref=\arabic{enumi}.\arabic*] 
			\item \label{step:ancilla-measurement-same} For all rounds in $\mathcal{M}_s$, Alice (Bob) first measures the system $A_c$ ($B_c$) in $\{\ket{0},\ket{1}\}$, learning her (his) choice of basis. If the result is $\ket{0}_{A_c}$ ($\ket{0}_{B_c}$), she (he) measures system $A$ ($B$) in $\{\ket{0},\ket{1}\}$; if the result is $\ket{1}_{A_c}$ ($\ket{1}_{B_c}$), she (he) measures system $A$ ($B$) in the Fock basis. 
			
			\item For all rounds in $\mathcal{M}_d$, Alice (Bob) measures the systems $A_c$ ($B_c$) and $A$ ($B$), using the same strategy as in step \ref{step:ancilla-measurement-same}.
			
			
		\end{enumerate}
		
		\item \textit{Intensity assignment}  \label{step:intensity-assignment}
		
		For all rounds in $\mathcal{M}$ in which Alice (Bob) obtained $\ket{1}_{A_c}$ ($\ket{1}_{B_c}$), she (he) assigns each $n$-photon state to intensity $\mu$ with probability $p_{\mu|n}$.
		
		\item \textit{Classical communication}
		
		For all rounds in $\mathcal{M}$, Alice and Bob announce all their basis and intensity choices over an authenticated public channel.
		
		\item \textit{Estimation of the number of phase errors}
		
		Alice and Bob calculate an upper bound on $N_{\textrm{ph}}$ using their $Z$ basis measurement data.
	\end{enumerate}
	\normalsize
	\noindent\rule{\textwidth}{1pt}
	
	Two points from the virtual protocol in Box 3 require further explanation. The first is that, in the real protocol, Bob flips his key bit when Eve reports $k_c = 0$ and $k_d = 1$. This step is omitted from the virtual protocol, since the $X$-basis bit flip gate $\sigma_z$ has no effect on Bob's $Z$-basis measurement result. The second point concerns step \ref{step:intensity-assignment}, which may appear to serve no purpose, but it is needed to ensure that the classical information exchanged between Alice and Bob is equivalent to that of the real protocol. The term $p_{\mu|n}$ is the probability that Alice's (Bob's) $Z$-basis $n$-photon pulse originated from intensity $\mu$, and it is given by
	\begin{equation}
	p_{\mu|n} = \frac{p_{\mu} p_{n|\mu}}{\sum_{\mu \in \underline{\mu}} p_{\mu} p_{n|\mu}}.
	\end{equation}

	\subsection*{Phase-error rate estimation}
	
	We now turn our attention to Alice and Bob's measurements in step (\ref{step:ancilla-measurement-same}) in Box 3. Let $u \in \{1,2,...,M_s\}$ index the rounds in $\mathcal{M}_s$, and let $\xi_u$ denote the measurement outcome of the $u$-th round. The possible outcomes are $\xi_u = X_{ij}$, corresponding to $\ket{00}_{A_c B_c}\ket{ij}_{A B}$, where $i,j \in \{0,1\}$; and $\xi_u = Z_{nm}$, corresponding to $\ket{11}_{A_c B_c}\ket{n,m}_{A B}$, where $n$ and $m$ are any positive integers. Note that the outcomes $\ket{10}_{A_c B_c}$ and $\ket{01}_{A_c B_c}$ are not possible due to the previous virtual sifting step. A phase error occurs when $\xi_u \in \{X_{00}, X_{11}\}$. In \cref{app:proof-conditional}, we prove that the probability to obtain a phase error in the $u$-th round, conditioned on all previous measurement outcomes in the protocol, is upper-bounded by
	\begin{equation} \label{eq:pherrorprob}
	\Pr \left(\xi_u \in \{X_{00}, X_{11}\}|\mathcal{F}_{u-1}\right) \leq \frac{p_X^2}{p_Z^2} \sum_{j=0}^{1} \Bigg[~ \sum_{n, m \in \mathds{N}_j} \sqrt{\frac{p_{nm|X}}{p_{nm|Z}} \Pr \left(\xi_u = Z_{n m}|\mathcal{F}_{u-1}\right)} ~\Bigg]^2,
	\end{equation}
	where $\mathcal{F}_{u-1}$ is the $\sigma$-algebra generated by random variables $\xi_1,...,\xi_{u-1}$, $\mathds{N}_0$ ($\mathds{N}_1$) is the set of non-negative even (odd) numbers, and the probability terms $p_{nm|X}$ and $p_{nm|Z}$ have been defined in \cref{eq:pnmX,eq:pnmZ}. In \cref{eq:pherrorprob}, for notation clarity, we have omitted the dependence of all probability terms on the outcomes of the measurements performed in steps (2) and (3) in Box 3.
	
	From \cref{eq:katoboundtrivial2}, we have that, except with probability $\varepsilon_a$,
	\begin{equation} \label{eq:azumaNph}
	\Nph \leq \sum_{u=1}^{M_s} 	\Pr \left(\xi_u \in \{X_{00}, X_{11}\}|\mathcal{F}_{u-1}\right) + \Delta,
	\end{equation}
	where $\Nph$ is the number of events of the form $\xi_u \in \{X_{00}, X_{11}\}$ in $\mathcal{M}_s$, and $\Delta = \sqrt{\frac{1}{2} M_s \ln \varepsilon_a^{-1}}$ is a deviation term. Similarly, from \cref{eq:katoboundtrivial2}, we have that, except with probability $\varepsilon_a$, 	
	\begin{equation} \label{eq:azumaMnm}
	\sum_{u=1}^{M_s} \Pr \left(\xi_u = Z_{nm}|\mathcal{F}_{u-1}\right) \leq M_{nm} + \Delta,
	\end{equation}
	where $M_{nm}$ is the number of events of the form $\xi_u = Z_{nm}$ in $\mathcal{M}_s$. As we will explain later, this bound is not tight when applied to the vacuum counts $M_{00}$. For this term, we use \cref{eq:katobound2}, according to which, except with probability $\varepsilon_a$, 
	
	\begin{equation} \label{eq:azumaM00}
	\sum_{u=1}^{M_s} \Pr \left(\xi_u = Z_{00}|\mathcal{F}_{u-1}\right) \leq M_{00} + \Delta_{00}.
	\end{equation}
	In this case, the deviation term is given by
	\begin{equation}
	\Delta_{00} = \left[b+a \left(\frac{2 M_{00}}{n} - 1 \right) \right] \sqrt{n},
	\end{equation}
	where $a$ and $b$ can be found by substituting $\tilde{\Lambda}_n$ by $\tilde{M}_{00}^U$ in \cref{eq:linearprogram2}.

	Now we will transform \cref{eq:pherrorprob} to apply \cref{eq:azumaNph,eq:azumaMnm,eq:azumaM00}. Let us denote the right-hand side of \cref{eq:pherrorprob} as $f(\vec{p}_u)$, where $\vec{p}_u$ is a vector of probabilities composed of $\Pr (\xi_u = Z_{n m}|\mathcal{F}_{u-1})$ $\forall n,m$. If we expand the square in $f(\vec{p}_u)$, we can see that all addends are positive and proportional to $\sqrt{p_1 p_2}$, where $p_1$ and $p_2$ are elements of $\vec{p}_u$, implying that $f(\vec{p}_u)$ is a concave function. Thus, by Jensen's inequality \cite{jensen1906fonctions}, we have
	\begin{equation} \label{eq:jensensinequality}
	\frac{1}{M_s} \sum_{u=1}^{M_s} f(\vec{p}_u) \leq  f\Bigg(\frac{1}{M_s} \sum_{u=1}^{M_s} \vec{p}_u\Bigg). 
	\end{equation}
	
	After taking the average over all rounds $M_s$ on both sides of \cref{eq:pherrorprob}, applying \cref{eq:jensensinequality} on the right-hand side, and cancelling out the term $1/M_s$ on both sides of the inequality, we have that
	\begin{equation} \label{eq:sum-pherrorprob}
	\begin{aligned}
	\sum_{u=1}^{M_s} \Pr \left(\xi_u \in \{X_{00}, X_{11}\}|\mathcal{F}_{u-1}\right) &\leq \frac{p_X^2}{p_Z^2} \sum_{j=0}^{1} \Bigg[~ \sum_{n,m \in \mathds{N}_j} \sqrt{\frac{p_{nm|X}}{p_{nm|Z}}  \sum_{u=1}^{M_s} \Pr \left(\xi_u = Z_{n m}|\mathcal{F}_{u-1}\right)}~\Bigg]^2.
	\end{aligned}
	\end{equation}
	
	We are now ready to apply \cref{eq:azumaNph,eq:azumaMnm,eq:azumaM00} to substitute the sums of probabilities by $N_{\textrm{ph}}$ and $M_{nm}$. However, note that Alice and Bob only estimate the value of $M_{n m}$ for terms of the form $n + m \leq \Scut$ and it is only useful to substitute \cref{eq:azumaMnm} for these terms. With this in mind, we obtain
	\begin{equation} \label{eq:Nph-intermediate}
	\begin{aligned}
	\Nph - \Delta \leq \frac{p_X^2}{p_Z^2} \sum_{j=0}^{1} \Bigg[\sum_{\substack{n,m \in \mathds{N}_j \\ n+m \leq \Scut}}  \sqrt{\frac{p_{nm|X}}{p_{nm|Z}}}\sqrt{M_{nm} + \Delta_{nm}} + \sum_{\substack{n,m \in \mathds{N}_j \\ n+m > \Scut}} \sqrt{\frac{p_{nm|X}}{p_{nm|Z}}  \sum_{u=1}^{M_s} \Pr \left(\xi_u = Z_{n m}|\mathcal{F}_{u-1}\right)} ~\Bigg]^2,
	\end{aligned}
	\end{equation}
	where $\Delta_{nm} = \Delta$ except for $\Delta_{00}$.
	
	We still need to deal with the sum over the infinitely many remaining terms of the form $n + m > \Scut$. For them, we apply the following upper bound
	\begin{equation} \label{eq:trivial_bound}
	\sum_{u=1}^{M_s} \Pr\left(\xi_u = Z_{n m}|\mathcal{F}_{u-1}\right) \leq  \sum_{u=1}^{M_s} \Pr\left(\xi_u = Z|\mathcal{F}_{u-1}\right) \leq M_Z + \Delta,
	\end{equation}
	where $\xi_u = Z$ denotes that Alice and Bob learn that they have used the $Z$ basis in the $u$th round in $\mathcal{M}_s$; and $M_Z$ is the number of events of the form $\xi_u = Z$ obtained by Alice and Bob. In the last step, we have used \cref{eq:katoboundtrivial2}, using an identical argument as in \cref{eq:azumaNph}. When we apply \cref{eq:trivial_bound} to \cref{eq:Nph-intermediate}, we end up with the term 
	\begin{equation}
	\label{eq:trivial_bound2}
	\sum_{\substack{n,m \in \mathds{N}_j \\ n+m > \Scut}} \sqrt{\frac{p_{nm|X}}{p_{nm|Z}}}\sqrt{M_{Z} + \Delta} = \sqrt{M_Z + \Delta} \sum_{\substack{n,m \in \mathds{N}_j \\ n+m > \Scut}}  \sqrt{\frac{p_{nm|X}}{p_{nm|Z}}}.
	\end{equation}
	We can show that \cref{eq:trivial_bound2} converges to a finite value if
	\begin{equation}
	\max \{\underline{\bf \mu}\} > \alpha^2.
	\end{equation}
	Substituting \cref{eq:trivial_bound} into \cref{eq:Nph-intermediate}, and isolating $N_{\textrm{ph}}$, we obtain
	\begin{equation}
	\label{eq:Nph-Mnm}
	N_{\textrm{ph}} \leq \frac{p_X^2}{p_Z^2} \sum_{j=0}^{1} \Bigg[ \sum_{\substack{n,m \in \mathds{N}_j \\ n+m \leq \Scut}}  \sqrt{\frac{p_{nm|X}}{p_{nm|Z}}}\sqrt{M_{n m} + \Delta_{nm}} + \sqrt{M_Z + \Delta} \sum_{\substack{n,m \in \mathds{N}_j \\ n+m > \Scut}}  \sqrt{\frac{p_{nm|X}}{p_{nm|Z}}} ~\Bigg]^2  + \Delta.
	\end{equation}
	
	Note that the right hand side of \cref{eq:Nph-Mnm} is a function of the measurement counts $M_{nm}$, which are unknown to the users. They must be substituted by the upper bounds $M_{nm}^U$ obtained via the decoy-state analysis. After doing so, we obtain \cref{eq:NphU}. The failure probability $\varepsilon$ associated to the estimation of $N_{\textrm{ph}}$ is upper-bounded by summing the failure probabilities of all concentration inequalities used. That includes each application of \cref{eq:katoboundtrivial2,eq:katobound2}, which fail with probability $\varepsilon_a$; and each application of the multiplicative Chernoff bound, which fails with probability $ \varepsilon_c$. In the case of three decoy intensities and $\Scut=4$, we have $\varepsilon = 9 \varepsilon_c + 10 \varepsilon_a$. In our simulations, we set $\varepsilon_c = \varepsilon_a$ for simplicity.

	\subsection*{Decoy-state analysis}
	
	Since Alice and Bob's $Z$-basis emissions are a mixture of Fock states, the measurement counts $M_{nm}$ have a fixed value, which is nevertheless unknown to them. Instead, the users have access to the measurement counts $M^{\mu \nu}$, the number of rounds in $\mathcal{M}_Z$ in which they selected intensities $\mu$ and $\nu$, respectively. To bound $M_{nm}$, we use the decoy-state method \cite{hwang2003quantum,lo2005decoy,wang2005beating}. This technique exploits the fact that Alice and Bob could have run an equivalent virtual scenario in which they directly send Fock states $\ket{n,m}$ with probability $p_{nm|Z}$. Then, after Eve's attack, they know the number $M_{nm}$ of successful events in which they respectively sent $n$ and $m$ photons, and they assign each of them to intensities $\mu$ and $\nu$ with probability
	\begin{equation}
	p_{\mu\nu|nm} = \frac{p_{\mu \nu} p_{nm|\mu \nu}}{p_{nm|Z}},
	\end{equation}
	where $p_{\mu \nu} = p_{\mu}p_{\nu}$ and $p_{nm|\mu\nu} = p_{n|\mu} p_{m|\nu}$. Each of these assignments can be regarded as an independent Bernoulli random variable, and the expected number of events $M_{\mu \nu}$ assigned to intensities $\mu$ and $\nu$ is 
	\begin{equation}
	\label{eq:M_ab_avg}
	\mathbb{E}[M^{\mu \nu}] = \sum_{n,m=0}^\infty p_{\mu \nu|nm} M_{nm}.
	\end{equation}
	In the actual protocol, Alice and Bob know the realisations $M^{\mu \nu}$ of these random variables. By using the inverse multiplicative Chernoff bound \cite{zhang2017improved,bahrani2019wavelength}, stated in \cref{app:chernoff}, they can compute lower and upper bounds $\mathbb{E}^L[M^{\mu \nu}]$ and $\mathbb{E}^{\text{U}}[M^{\mu \nu}]$ for $\mathbb{E}[M^{\mu \nu}]$. These will set constraints on the possible value of the terms $M_{nm}$. We are interested in the indices $(i,j)$ such that $i+j \leq \Scut$ and $i+j$ is even, and an upper bound on each $M_{ij}$ can be found by solving the following linear optimisation problem
	\begin{equation}
	\label{eq:linearprogram}
	\begin{aligned}
	&\max M_{ij}  \\
	&\textrm{s.t. } \forall \mu,\nu & \mathbb{E}^{\text{U}}[M^{\mu \nu}] \geq \sum_{n,m=0}^\infty p_{\mu \nu|nm} M_{nm}, \\
	&& \mathbb{E}^L[M^{\mu \nu}] \leq \sum_{n,m=0}^\infty p_{\mu \nu|nm} M_{nm}.
	\end{aligned}
	\end{equation}
	
	This problem can be solved numerically using linear programming techniques, as described in the Supplementary Note 2 of \cite{curty2014finite}. While accurate, this method can be computationally demanding. For this reason, we have instead adapted the asymptotic analytical bounds of \cite{grasselli2019practical,FedAlv} to the finite-key scenario and used them in our simulations. The results obtained using these analytical bounds are very close to those achieved by numerically solving \cref{eq:linearprogram}. This analytical method is described in \cref{app:decoy-analysis}.

	\subsection*{Concentration inequality for sums of dependent random variables}
	
	A crucial step in our analysis is the substitution of the sums of probabilities in \cref{eq:sum-pherrorprob} by their actual realisation in the protocol. Typically, this is done by applying the well-known Azuma's inequality \cite{azuma1967weighted}. Instead, we use the following novel result \cite{kato2020bound}:
	
	Let $\xi_1,..., \xi_n$ be a sequence of random variables satisfying $0\leq \xi_l \leq 1$, and let $\Lambda_l = \sum_{u=1}^{l} \xi_u$. Let $\mathcal{F}_{l}$ be its natural filtration, i.e. the $\sigma$-algebra generated by $\{\xi_1,...,\xi_l\}$. For any $n$, and any $a,b$ such that $b\geq \abs{a}$,
    %
    %
        \begin{equation}
    \label{eq:katobound}
        \Pr \left[\sum_{u=1}^{n} E(\xi_u \vert \mathcal{F}_{u-1}) - \Lambda_n  \geq \left[b+a \left(\frac{2 \Lambda_n}{n} - 1 \right) \right] \sqrt{n} \right] \leq \exp \left[\frac{-2(b^2-a^2)}{(1+\frac{4a}{3 \sqrt{n}})^2}  \right].
    \end{equation}
    By replacing $\xi_l \to 1-\xi_l$ and $a \to -a$, we also derive
    
     \begin{equation}
        \label{eq:katoboundinv}
        \Pr \left[\Lambda_n - \sum_{u=1}^{n} E(\xi_u \vert \mathcal{F}_{u-1})  \geq \left[b+a \left(\frac{2 \Lambda_n}{n} - 1 \right) \right] \sqrt{n} \right] \leq \exp \left[\frac{-2(b^2-a^2)}{(1-\frac{4a}{3 \sqrt{n}})^2} \right].
    \end{equation}  
    In our analysis, we apply \cref{eq:katobound,eq:katoboundinv} to sequences $\xi_1,..., \xi_n$ of Bernoulli random variables, for which $E(\xi_u \vert \mathcal{F}_{u-1}) = \Pr(\xi_u = 1 \vert \mathcal{F}_{u-1})$.
    
    Now, if we set $a = 0$ on \cref{eq:katobound,eq:katoboundinv}, we obtain
    
    \begin{equation}
    \label{eq:katoboundtrivial}
    \begin{gathered}
         \Pr \left[\Lambda_n - \sum_{u=1}^{n} \Pr(\xi_u = 1 \vert \mathcal{F}_{u-1}) \geq b \sqrt{n} \right] \leq \exp \left[-2 b^2 \right],\\  
         \Pr \left[\sum_{u=1}^{n} \Pr(\xi_u = 1 \vert \mathcal{F}_{u-1}) - \Lambda_n \geq b \sqrt{n} \right] \leq \exp \left[-2 b^2 \right].
    \end{gathered}
    \end{equation}
    This is a slightly improved version of the original Azuma's inequality, whose the right-hand side is $\exp \left[-\frac{1}{2} b^2 \right]$. Equating the right hand sides of \cref{eq:katoboundtrivial} to $\varepsilon_a$, and solving for $b$, we have that
    
    \begin{equation}
    \label{eq:katoboundtrivial2}    
    \begin{gathered}
	\sum_{u=1}^{n} \Pr \left(\xi_u = 1 \vert\xi_1,..., \xi_{u-1}\right) \leq \Lambda_n  + \Delta, \\
	\Lambda_n \leq  \sum_{u=1}^{n} \Pr \left(\xi_u = 1 \vert\xi_1,..., \xi_{u-1}\right) + \Delta,
	\end{gathered}
    \end{equation}
    with $\Delta = \sqrt{\frac{1}{2} n \ln \varepsilon_a^{-1}}$, and where each of the bounds in \cref{eq:katoboundtrivial2} fail with probability at most $\varepsilon_a$. 
    
    The bound in \cref{eq:katoboundtrivial2} scales with $\sqrt{n}$, and it is only tight when $\Lambda_n$ is of comparable magnitude to $n$. When $\Lambda_n \ll n$, one can set $a$ and $b$ in \cref{eq:katobound} appropriately to obtain a much tighter bound. To do so, one can use previous knowledge about the channel to come up with a prediction $\tilde{\Lambda}_n$ of $\Lambda_n$ before running the experiment. Then, one obtains the values of $a$ and $b$ that would minimise the deviation term if the realisation of $\Lambda_n$ equalled $\tilde{\Lambda}_n$, by solving the optimisation problem
    
    \begin{equation}
    \label{eq:linearprogram2}
    \begin{aligned}
    &\min &  \left[b+a \left(\frac{2 \tilde{\Lambda}_n}{n} - 1 \right) \right] \sqrt{n} \\
    &\textrm{s.t. }  &\exp \left[\frac{-2(b^2-a^2)}{(1+\frac{4a}{3 \sqrt{n}})^2} \right] = \varepsilon_a, \\
    && b \geq \abs{a}.
    \end{aligned}
    \end{equation}
    The solution to \cref{eq:linearprogram2} is
    \begin{equation}
    \label{eq:ab-solution}
    \begin{gathered}
        a = \frac{3 \left(72 \sqrt{n} X  (n-X)\ln \varepsilon_a-16 n^{3/2} \ln^2\varepsilon_a+9 \sqrt{2} (n-2 X) \sqrt{-n^2 \ln \varepsilon_a  (9 X (n-X)-2 n \ln \varepsilon_a)}\right)}{4 (9 n-8\ln
   \varepsilon_a) (9 X (n-X)-2 n \ln\varepsilon_a)}, \\
   b = \frac{\sqrt{18 a^2 n-\left(16 a^2+24 a \sqrt{n}+9 n\right) \ln\varepsilon_a}}{3 \sqrt{2n}}.
   \end{gathered}
    \end{equation}
	After fixing $a$ and $b$, we have that
	
	\begin{equation}
	\label{eq:katobound2}
		\sum_{u=1}^{n} \Pr \left(\xi_u = 1 \vert\xi_1,..., \xi_{u-1}\right) \leq \Lambda_n  + \Delta',
	\end{equation}
	except with probability $\varepsilon_a$, where
	\begin{equation}
	\Delta' = \left[b+a \left(\frac{2 \Lambda_n}{n} - 1 \right) \right] \sqrt{n}.
	\end{equation}
	
	In our numerical simulations, we have found the simple bound in \cref{eq:katoboundtrivial2} to be sufficiently tight for all components except the vacuum contribution $M_{00}$. For this latter component, we use \cref{eq:katobound2} instead. However, note that the users do not know the true value of $M_{00}$, even after running the experiment. Instead, they will obtain an upper-bound $M_{00}^U$ on $M_{00}$ via the decoy-state method, and they will apply \cref{eq:katobound2} to this upper bound. Therefore, to optimise the bound, the users should come up with a prediction $\tilde{M}_{00}^U$ on the value of $M_{00}^U$ that they expect to obtain after running the experiment and performing the decoy-state analysis, and then substitute $ \tilde{\Lambda}_n \to \tilde{M}_{00}^U$ in \cref{eq:linearprogram2} to obtain the optimal values of $a$ and $b$. To find $\tilde{M}_{00}^U$, one can simply use their previous knowledge of the channel to come up with predictions $\tilde{M}^{\mu \nu}$ of $M^{\mu \nu}$, and run the decoy-state analysis using these values to obtain $\tilde{M}_{00}^U$.

	\section*{Acknowledgements}
	
    We thank Margarida Pereira, Kiyoshi Tamaki and Mirko Pittaluga for valuable discussions. We thank Kento Maeda, Toshihiko Sasaki and Masato Koashi for the computer code used to generate \cref{fig:graph4}, as well as for insightful discussions. This work was supported by the European Union's Horizon 2020 research and innovation programme under the Marie Sklodowska-Curie grant agreement number 675662 (QCALL). M.C. also acknowledges support from the Spanish Ministry of Economy and Competitiveness (MINECO), and the Fondo Europeo de Desarrollo Regional (FEDER) through the grant TEC2017-88243-R. K.A. thanks support, in part, from PRESTO, JST JPMJPR1861. A.N. acknowledges support from a FPU scholarship from the Spanish Ministry of Education. M.R. acknowledges the support of UK EPSRC Grant EP/M013472/1. G.K. acknowledges financial support by the JSPS Kakenhi (C) No. 17K05591.
	
	\section*{Author contributions}
    G.C.L. performed the analytical calculations and the numerical simulations. A.N. constructed the analytical decoy-state estimation method. All the authors contributed to discussing the main ideas of the security proof, checking the validity of the results, and writing the paper.
	
	\appendix
	
	\clearpage

	\section{Security bounds}
	\label{app:security-bounds}
	
	Let \textbf{X} (\textbf{X}$'$) denote Alice's (Bob's) sifted key of length $M_X$ before the post-processing step of the protocol. After the error correction and verification steps, Bob should have turned \textbf{X}$'$ into a copy of \textbf{X}. Then, Alice and Bob apply a privacy amplification scheme based on two-universal hashing to obtain a shorter secret key of length $\ell$. The protocol is $\epsilon_{\textrm{s}}$-secret if \cite{tomamichel2012tight}
	\begin{equation}
	\label{eq:epsilonsec}
	\epsilon_{\textrm{s}} \leq 	2 \varepsilon + \frac{1}{2} \sqrt{2^{\ell-H_{\min}^{\varepsilon}(\textbf{X} \vert E')}},
	\end{equation}
	where $E'$ represents Eve's total side information about \textbf{X}, and $H_{\min}^{\varepsilon}(\textbf{X} \vert E')$ is the $\varepsilon$-smooth min entropy of \textbf{X} conditioned on $E'$. Let $E$ denote Eve's side information before the error correction step. By the chain rule for smooth min-entropies \cite{tomamichel2012tight}, 
	\begin{equation}
	\label{eq:chainminentropy}
	H_{\min}^{\varepsilon}(\textbf{X} \vert E') \geq H_{\min}^{\varepsilon}(\textbf{X} \vert E) -  \lambda_{\textrm{EC}} - \log_2 \frac{2}{\epsilon_{\textrm{c}}},
	\end{equation}
	where $\lambda_{\textrm{EC}}$ ($\log_2 \frac{2}{\epsilon_{\textrm{c}}}$) is the number of bits revealed in the error correction (verification) step of the protocol. We now make use of the following theorem, introduced in \cite{tomamichel2011uncertainty}, which we reproduce here for completeness.
	
	\hfill
	
	\textbf{Theorem} \cite{tomamichel2011uncertainty}: Let $\varepsilon > 0$, $\rho_{AEB}$ be a tripartite quantum state,  $\mathds{X} = \{M_x\}$ and  $\mathds{Z} = \{N_z\}$ be two POVMs on $A$, and \textbf{X} (\textbf{Z}) be the result of the measurement of $\mathds{X}$ ($\mathds{Z}$). Then,
	\begin{equation}
	\label{eq:uncertaintytheorem}
	H_{\min}^{\varepsilon}(\textbf{X} \vert E) + H_{\max}^{\varepsilon}(\textbf{Z} \vert B)  \geq q,
	\end{equation}
	where $q = -\log_2 \frac{1}{c}$, with
	\begin{equation}
	\label{eq:q}
	c = \max_{x,z} \norm{\sqrt{M_x} \sqrt{N_z}}^2_{\infty}.
	\end{equation}
	
	To apply this theorem, we consider a slight modification to our virtual protocol in Box 3. In step (\ref{step:ancilla-measurement-same}), Alice and Bob now first measure all the basis ancillas $A_c$ and $B_c$ in $\mathcal{M}_s$, keeping only the $M_X$ successful rounds in which they used the $X$ basis, which we denote as the set $\mathcal{M}_X$. Let $\rho_{AEB}$ be the quantum state that describes all systems $A$ and $B$ in $\mathcal{M}_X$, as well as Eve's side information $E$ on them. Note that if Alice measures all her systems $A$ in the $X$ basis, she will obtain a raw key \textbf{X} that is identical to the one she would have obtained in the real protocol; while if she measures them in the $Z$ basis, she will obtain a raw key \textbf{Z} that is identical to that of the virtual protocol.
	
	Let $\mathds{X} = \{M_x\}$ ($\mathds{Z} = \{N_z\}$) denote Alice's overall POVM if she chooses to measure all her systems $A$ in the $X$ ($Z$) basis. The elements $M_x$ of $\mathds{X}$ are of the form $\ketbra{x_1 x_2 x_3 ...}$, where $x_n \in \{+,-\}$ is the result of the measurement of round $n \in \{1,...,M_X\}$. Conversely, the elements $N_z$ of $\mathds{Z}$ are of the form  $\ketbra{z_1 z_2 z_3 ...}$, with $z_n \in \{0,1\}$. Since $M_x$ and $N_z$ are rank 1 projective measurements, we have that 
	\begin{equation}
	\max_{x,z} \norm{\sqrt{M_X} \sqrt{N_z}}^2_{\infty} = \max_{x,z} \norm{\braket{x_1 x_2 x_3 ...}{z_1 z_2 z_3 ...}}^2 = 2^{-M_X},
	\end{equation}
	where in the last step we have used the fact that $\norm{\braket{x_n}{z_n}}^2 = 1/2$, independently of the value of $x_n$ and $z_n$. From  \cref{eq:uncertaintytheorem}, it follows that
	\begin{equation}
	H_{\min}^{\varepsilon}(\textbf{X} \vert E) + H_{\max}^{\varepsilon}(\textbf{Z} \vert B)  \geq M_X.
	\end{equation}
	
	Now, let us assume that Bob measures his systems $B$ using POVM $\mathds{Z}$, obtaining a string \textbf{Z}$'$ that is identical to the one that he would obtain in the virtual protocol. Clearly, the result of a measurement of $B$ cannot contain more information about \textbf{Z} than system $B$ itself, and therefore
	\begin{equation}
	H_{\max}^{\varepsilon}(\textbf{Z} \vert \textbf{Z}') \geq H_{\max}^{\varepsilon}(\textbf{Z} \vert B),
	\end{equation}
	from which we finally obtain
	\begin{equation}
	\label{eq:uncertaintyrelation}
	H_{\min}^{\varepsilon}(\textbf{X} \vert E) + H_{\max}^{\varepsilon}(\textbf{Z} \vert \textbf{Z}')  \geq M_X.
	\end{equation}
	
	In the Methods section, we show that the error rate between \textbf{Z} and \textbf{Z}$'$ is bounded by $e_{\textrm{ph}}^{\text{U}}$, except with a failure probability $\varepsilon$. Therefore \cite{tomamichel2012tight},
	\begin{equation}
	\label{eq:maxentropyandpherror}
	H_{\max}^{\varepsilon}(\textbf{Z} \vert \textbf{Z}') \leq M_X h(e_{\textrm{ph}}^{\text{U}}),
	\end{equation}
	and by combining \cref{eq:chainminentropy,eq:uncertaintyrelation,eq:maxentropyandpherror}, we have that
	\begin{equation}
	\label{eq:hmin}	
	H_{\min}^{\varepsilon}(\textbf{X} \vert E') \leq M_X \left[1-h(e_{\textrm{ph}}^{\text{U}})\right] - \lambda_{\textrm{EC}} - \log_2 \frac{2}{\epsilon_{\textrm{c}}}.
	\end{equation}
	
	By substituting \cref{eq:hmin} and the secret key length $\ell$ given in \cref{eq:keyrate} of the main text into \cref{eq:epsilonsec}, we finally obtain that the protocol is $\epsilon_{\textrm{s}}$-secret if we choose $\epsilon_{\textrm{s}}$ such that
	\begin{equation}
	\epsilon_{\textrm{s}} \leq 2 \varepsilon + \epsilon_{\textrm{PA}}.
	\end{equation}
	
	\section{Analytical estimation method}
	\label{app:decoy-analysis}
	In this Note, we present an analytical method to obtain the upper bounds $M_{nm}^U$ in \cref{eq:NphU}, using the observed quantities $M^{\mu\nu}$. First, we explain the general idea behind the procedure, and then we obtain specific analytical bounds for the case of three decoy intensities and $\Scut = 4$, which we use in our simulations. We have numerically verified that the choice of three decoy intensities is optimal for reasonable block size values below $10^{12}$ transmitted signals. 
	
	Our starting point is \cref{eq:M_ab_avg}, which we rewrite as
	%
	%
	\begin{equation}\label{Eq_M_abhat}
	\hat{M}^{\mu\nu}=\sum_{n,m=0}^{\infty}\frac{\mu^n\nu^m}{n!m!p_{n|Z}p_{m|Z}}M_{nm},
	\end{equation}
	by defining $\hat{M}^{\mu\nu}=e^{\mu+\nu}\frac{\mathbb{E}[M^{\mu\nu}]}{p_{\mu}p_{\nu}}$. To obtain an upper bound for a specific term $M_{ij}$ in \cref{Eq_M_abhat}, we follow a procedure analogous to Gaussian elimination, defining a linear combination
	\begin{equation}\label{eq:M_mu}
	    \Omega = \sum_{\mu,\nu} \hat{c}_{\mu \nu} \hat{M}^{\mu \nu}.
	\end{equation}
	From \cref{Eq_M_abhat}, $\Omega$ can also be expressed as a linear combination of the $M_{nm}$ terms, that is,
	\begin{equation}\label{eq:Fij}
    \Omega = \sum_{n,m=0}^{\infty} c_{nm} M^{nm}.
	\end{equation}
	%
	Then, we rewrite the R.H.S. of \cref{eq:Fij} as
	\begin{equation}\label{eq:Fij_div}
	\Omega=c_{ij}M_{ij}+\sum_{(n,m)\in S_+}c_{nm}M_{nm}+\sum_{(n,m)\in S_-}c_{nm}M_{nm},
	\end{equation}
	where we have singled out the index $(i,j)$, ensured that $c_{ij} >0$, and defined $S_+$ ($S_-$) as the set of pairs $(n,m)\neq(i,j)$ such that $c_{nm}$ is a positive (negative) number. From \cref{eq:Fij_div}, one can obtain the following upper bound on $M_{ij}$
	\begin{align}\label{eq:Mij_bound}
	M_{ij}&=\frac{1}{c_{ij}}\left[\Omega-\sum_{nm\in S_+}c_{nm}M_{nm}-\sum_{nm\in S_-}c_{nm}M_{nm}\right]\nonumber\\
	&\leq\frac{1}{c_{ij}}\left[\text{UB}\{\Omega\}-\text{LB}\{\sum_{(n,m)\in S_+}c_{nm}M_{nm}\}+\text{UB}\{\sum_{(n,m)\in S_-}|c_{nm}|M_{nm}\}\right],
	\end{align}
	where UB$\{x\}$ (LB$\{x\}$) denotes an upper (lower) bound on $x$.
	
	Now, we find an expression for each of the bounds within \cref{eq:Mij_bound}. First, note that $\Omega$ is a linear combination of the expected values $\mathbb{E}[M^{\mu\nu}]$ as in Eq.~(\ref{Eq_M_abhat}). While these are unknown to the users, they can obtain lower and upper bounds $\mathbb{E}^{\rm L}[M^{\mu\nu}]$ and $\mathbb{E}^{\rm U}[M^{\mu\nu}]$ using the inverse multiplicative Chernoff bound presented in \cref{app:chernoff}. To obtain $\text{UB}\{\Omega\}$, we simply replace each term $\mathbb{E}[M^{\mu\nu}]$ in Eq.~(\ref{eq:M_mu}) by either its upper or lower bound, depending on whether its coefficient in $\Omega$ is positive or negative.
	
	Second, we use the fact that $M_{nm} \geq 0$ to find the trivial lower bound $\text{LB}\{\sum_{(n,m)\in S_+}c_{nm}M_{nm}\} = 0$. For the remaining term, we note that
	\begin{align}\label{eq:UB_NegativeSet}
	\sum_{n,m\in S_-}|c_{nm}|M_{nm}&\leq c_{\max}\sum_{n,m\in S_-}M_{nm}\nonumber\\
	&\leq c_{\max}(M_{z}-M_{ij}-\text{LB}\{\sum_{\substack{n,m\notin S_-\\(n,m)\neq (i,j)}}M_{nm}\}),
	\end{align}
	where we have chosen $c_{\rm max}$ such that $c_{\rm max}\geq|c_{nm}|$ for all the pairs $(n,m) \in {\cal S}_-$, and the last lower bound depends on the particular $M_{ij}$ that we are trying to estimate, as we will show later. Substituting the three bounds in \cref{eq:Mij_bound} and isolating $M_{ij}$, we obtain
	\begin{align}
	M_{ij}&\leq\frac{1}{c_{ij}+c_{\max}}\left[\text{UB}\{\Omega\}+c_{\max}(M_{z}-\text{LB}\{\sum_{\substack{n,m\notin S_-\\(n,m)\neq (i,j)}}M_{nm}\})\right].
	\end{align}

	\subsection{Bounds for three decoy intensities}
	Now, we obtain explicit lower bounds for the case in which $\Scut = 4$ and each of Alice and Bob use three different intensity settings, satisfying $\mu_0>\mu_1>\mu_2$ and $\nu_0>\nu_1>\nu_2$, respectively. For this, we take inspiration from the asymptotic analytical bounds derived in~\cite{FedAlv}. First, we define
	\begin{equation}\label{eq:k_general}
	K^{a_S,a_I,b_S,b_I}= 
	\kappa^{a_I}_A\kappa^{b_I}_B\hat{M}^{a_S,b_S} - \kappa^{a_S}_A\kappa^{b_I}_B\hat{M}^{a_I,b_S} - \kappa^{a_I}_A\kappa^{b_S}_B\hat{M}^{a_S,b_I} + \kappa^{a_S}_A\kappa^{b_S}_B\hat{M}^{a_I,b_I},
	\end{equation}
	which is a function of some intensities that satisfy $a_S>a_I$ and $b_S>b_I$, with $a_S,a_I\in\{\mu_0,\mu_1,\mu_2\}$ and $b_S,b_I\in\{\nu_0,\nu_1,\nu_2\}$. The coefficients $\kappa^{\mu}_A$ and $\kappa^{\nu}_B$ depend on the specific $M_{ij}$ that is to be estimated, but we have omitted this dependence from the notation for simplicity. Using the previous equation, we now define
	\begin{align}\label{eq:omega_general}
	\Omega&= 
	w^{\mu_1\mu_2}_Aw^{\nu_1\nu_2}_BK^{\mu_0,\mu_1,\nu_0,\nu_1}-
	w^{\mu_0\mu_1}_Aw^{\nu_1\nu_2}_BK^{\mu_1,\mu_2,\nu_0,\nu_1}\nonumber\\
	&-w^{\mu_1\mu_2}_Aw^{\nu_0\nu_1}_BK^{\mu_0,\mu_1,\nu_1,\nu_2}+
	w^{\mu_0\mu_1}_Aw^{\nu_0\nu_1}_BK^{\mu_1,\mu_2,\nu_1,\nu_2},
	\end{align}
	where the coefficients $w^{\mu\nu}_A$ and $w^{\mu\nu}_B$ also depend on the particular $M_{ij}$ that we want to estimate. If we rewrite $\Omega$ as  $\Omega=\sum_{k,l=0}^{2}\hat{c}_{\mu_k\nu_l}\hat{M}^{\mu_k\nu_l}$, it is easy to prove that if the coefficients $w^{\mu\nu}_A$, $w^{\mu\nu}_B$, $\kappa^{\mu}_A$ and $\kappa^{\mu}_B$ are all positive, the coefficients $\hat{c}_{\mu_k\nu_l}$ are always positive (negative) when $k+l$ is even (odd). Thus, one can find upper ($\Omega^{\rm U}$) and lower ($\Omega^{\rm L}$) bounds on $\Omega$ by properly replacing each $\hat{M}_{\mu_k\nu_l}$ by either its upper or lower bound, as explained in the introduction of this Note.

	\subsubsection{Upper bound on \texorpdfstring{$M_{00}$}{M00}}
	By substituting $\kappa^{\mu}_A=\kappa^{\mu}_B=\mu$ and $w^{\mu\nu}_A=w^{\mu\nu}_B=(\mu^2\nu-\nu^2\mu)$ in $\Omega$, we obtain the following function $\Omega_{00}$:
	\begin{equation}
	\Omega_{00}:= \Omega = c_{00}M_{00}+\sum_{n=3}^{\infty}c_{n0}M_{n0}+\sum_{m=3}^{\infty}c_{0m}M_{0m}+\sum_{\substack{n=3\\m=3}}^{\infty}c_{nm}M_{nm},
	\end{equation}
	where the coefficients
	\begin{eqnarray}
	c_{nm}&=&\frac{1}{m! n! p_{nm|Z}}\mu_1 \nu_1 \left[\mu_0 \mu_1 \left(\mu_0-\mu_1\right) \mu_2^n- \mu_0\mu_2 \left(\mu_0-\mu_2\right) \mu_1^n+\mu_1\mu_2 \left(\mu_1-\mu_2\right) \mu_0^n\right]\nonumber\\
	&&\times\left[\nu_0 \nu_1 \left(\nu_0-\nu_1\right) \nu_2^m-\nu_0\nu_2 \left(\nu_0-\nu_2\right) \nu_1^m+\nu_1\nu_2 \left(\nu_1-\nu_2\right) \nu_0^m\right],
	\end{eqnarray}
	can be shown to be non-negative for all $n,m$~\cite{FedAlv}. Then, an upper bound on $M_{00}$ is straightforwardly given by
	\begin{equation}
	M_{00}\leq \frac{\Omega_{00}^{\text{U}}}{c_{00}},
	\end{equation}
	where we have lower bounded the term $\sum_{n=3}^{\infty}c_{n0}M_{n0}+\sum_{m=3}^{\infty}c_{0m}M_{0m}+\sum_{\substack{n=3\\m=3}}^{\infty}c_{nm}M_{nm}$ by zero since all the coefficients satisfy $c_{nm}\geq 0$.

	\subsubsection{Upper bound on  \texorpdfstring{$M_{11}$}{M11}}
	By substituting $\kappa^{\mu}_A=\kappa^{\mu}_B=1$ and $w^{\mu\nu}_A=w^{\mu\nu}_B=(\mu^2-\nu^2)$ in $\Omega$, we obtain the following function $\Omega_{11}$:
	\begin{equation}\label{eq:M11sum}
	\Omega_{11}:= \Omega = c_{11}M_{11}+\sum_{n=3}^{\infty}c_{n1}M_{n1}+\sum_{m=3}^{\infty}c_{1m}M_{1m}+\sum_{\substack{n=3\\m=3}}^{\infty}c_{nm}M_{nm},
	\end{equation}
	where the coefficients
	\begin{eqnarray}
	c_{nm}&=&\frac{\left[ \mu_0^n\left(\mu_1^2-\mu_2^2\right)-\mu_1^n \left(\mu_0^2-\mu_2^2\right)+\mu_2^n \left(\mu_0^2-\mu_1^2\right)\right] \left[\nu_0^m\left(\nu_1^2-\nu_2^2\right)-\nu_1^m \left(\nu_0^2-\nu_2^2\right)+\nu_2^m \left(\nu_0^2-\nu_1^2\right)\right]}{m! n! p_{nm|Z}},\label{eq:cnm11}
	\end{eqnarray}
	can be shown to be negative for the pairs $(n,m)\in S_-$, with $S_-=\{(n,m)|n\geq3,m=1\}\cup\{(n,m)|n=1,m\geq3\}$ and non-negative for the paris $(n,m)\in S_{+}$, with $S_+=\{(n,m)|n\geq3,m\geq3\}$~\cite{FedAlv}. According to Eq.~(\ref{eq:UB_NegativeSet}), an upper bound on the sum of negative terms can be obtained by
	\begin{eqnarray}\label{eq:M11_bound_negative}
	\sum_{n,m\in S_-}|c_{nm}|M_{nm}&\leq& c_{\max}(M_{z}-M_{11}-\text{LB}\{\sum_{\substack{n,m\notin S_-\\(n,m)\neq (1,1)}}M_{nm}\})\nonumber\\
	&\leq&c_{\max}(M_{z}-M_{11}-\text{LB}\{\sum_{n,m\in S_{o}}M_{nm}\})\nonumber\\
	&\leq&c_{\max}(M_{z}-M_{11}-\text{LB}\{\sum_{n=0}^{\infty}M_{n0}\}-\text{LB}\{\sum_{m=0}^{\infty}M_{0m}\}+\text{UB}\{M_{00}\}),
	\end{eqnarray}
	where $c_{\max}\geq |c_{nm}|$ for all the pairs $(n,m)\in S_-$  and $S_{o}=\{(n,0)|n\geq0\}\cup\{(0,m)|m\geq0\}$. In Eq.~(\ref{eq:M11_bound_negative}), the second inequality comes from the fact that we have set to zero all those terms $M_{nm}$, with $(m,n)\neq(1,1)$, which do not belong to $S_{-}$ nor to $S_{o}$ because $M_{nm}\geq 0$, $\forall n,m$. A valid $c_{\max}$ can be obtained by noticing that, for $n>s$,
	\begin{eqnarray}
	g_{\mu}(n)&:=&\frac{ \mu_0^n\left(\mu_1^s-\mu_2^s\right)-\mu_1^n \left(\mu_0^s-\mu_2^s\right)+\mu_2^n \left(\mu_0^s-\mu_1^s\right)}{n! p_{n|Z}}\nonumber\\
	&\leq&\frac{ \mu_0^n\left(\mu_1^s-\mu_2^s\right)}{n! p_{n|Z}}\nonumber\\
	&\leq&\frac{\left(\mu_1^s-\mu_2^s\right)}{e^{-\mu_0}p_{\mu_0}}.\label{eq:cmax}
	\end{eqnarray}
	This means that, from Eqs.~(\ref{eq:cnm11}) and (\ref{eq:cmax}), $c_{\max}$ is given by
	\begin{equation}
	c_{\max}=\max[\frac{\left(\mu_1^2-\mu_2^2\right)}{e^{-\mu_0}p_{\mu_0}}|g_{\nu}(1)|,\frac{\left(\nu_1^2-\nu_2^2\right)}{e^{-\nu_0}p_{\nu_0}}|g_{\mu}(1)|].
	\end{equation}
	Finally from Eqs.~(\ref{eq:Mij_bound}) and (\ref{eq:M11_bound_negative}), an upper bound on $M_{11}$ is given by
	\begin{equation}\label{eq:M11}
	M_{11}\leq M^{\text{U}}_{11}= \frac{\Omega_{11}^{\text{U}}+c_{\max}(M_{z}-M_{0A}^L-M_{0B}^L+M_{00}^{\text{U}})}{c_{11}+c_{\max}},
	\end{equation}
	where $M_{0A}^L$ and $M_{0B}^L$ are lower bounds on the quantities $M_{0A}=\sum_{m=0}^{\infty}M_{0m}$ and $M_{0B}=\sum_{n=0}^{\infty}M_{n0}$, respectively, and we have lower bounded the term $\sum_{\substack{n,m=3}}^{\infty}c_{nm}M_{nm}$ by zero. Since $M_{0A}$ and $M_{0B}$ depend only on a single emitter, we can estimate them using the same method as for the vacuum component in BB84. Using the results of \cite{lim2014concise}, we have that
	
	\begin{align}
	M_{0A}^L&=p_0\frac{\mu_1\text{LB}\{\hat{M}^{\mu_2}\}-\mu_2\text{UB}\{\hat{M}^{\mu_1}\}}{\mu_1-\mu_2},\label{eq:M0AL}\\
	M_{0B}^L&=p_0\frac{\nu_1\text{LB}\{\hat{M}^{\nu_2}\}-\nu_2\text{UB}\{\hat{M}^{\nu_1}\}}{\nu_1-\nu_2},\label{eq:M0BL}
	\end{align}
	where $\hat{M}^{\mu}=e^{\mu}\frac{\mathbb{E}[M^{\mu}]}{p_{\mu}}$, $\hat{M}^{\nu}=e^{\nu}\frac{\mathbb{E}[M^{\nu}]}{p_{\nu}}$, $M^{\mu}=\sum_{\nu}M^{\mu\nu}$ and $M^{\nu}=\sum_{\mu}M^{\mu\nu}$, with $\mu\in\{\mu_0,\mu_1,\mu_2\}$ and $\nu\in\{\nu_0,\nu_1,\nu_2\}$; and the upper and lower bounds included in Eqs.~(\ref{eq:M0AL}) and~(\ref{eq:M0BL}) are obtained accordingly to Eq.~(\ref{eq:chernoff-begin}).

	\subsubsection{Upper bound on  \texorpdfstring{$M_{22}$}{M22}}
	By substituting $\kappa^{\mu}_A=\kappa^{\mu}_B=1$ and $w^{\mu\nu}_A=w^{\mu\nu}_B=(\mu-\nu)$ in $\Omega$, we obtain the following function $\Omega_{22}$:
	\begin{equation}
	\Omega_{22}:= \Omega =\sum_{n,m=2}^{\infty}c_{nm}M_{nm},
	\end{equation}
	where the coefficients
	\begin{equation}
	c_{nm}=\frac{\left[\mu_0^n\left(\mu_1-\mu_2\right) - \mu_1^n\left(\mu_0-\mu_2\right)+\mu_2^n\left(\mu_0-\mu_1\right)\right] \left[\nu_0^m\left(\nu_1-\nu_2\right) - \nu_1^m\left(\nu_0-\nu_2\right)+\nu_2^m\left(\nu_0-\nu_1\right)\right]}{m! n! p_{nm|Z}},
	\end{equation}
	can be shown to be non-negative for all the pairs $(n,m)$~\cite{FedAlv}. Then, an upper bound on $M_{22}$ is straightforwardly given by
	\begin{equation}
	M_{22}\leq M_{22}^{\text{U}} = \frac{\Omega_{22}^{\text{U}}}{c_{22}},
	\end{equation}
	where we have lower bounded the term $\sum_{\substack{n,m\geq 2\\n,m\neq 2}}^{\infty}c_{nm}M_{nm}$ by zero.

	\subsubsection{Upper bounds on \texorpdfstring{$M_{02}$}{M02} and  \texorpdfstring{$M_{04}$}{M04}}
	By substituting $\kappa^{\mu}_A=\mu$, $\kappa^{\mu}_B=1$, $w^{\mu_0\mu_1}_A=(\mu_0-\mu_1)\mu_0$, $w^{\mu_1\mu_2}_A=(\mu_1-\mu_2)\mu_2$ and $w^{\mu\nu}_B=(\mu-\nu)$ in $\Omega$, we consider the following function $\Omega_{02}$:
	\begin{equation}\label{eq:M02sum}
	\Omega_{02}:= \Omega =\sum_{m=2}^{\infty}c_{0m}M_{0m}+\sum_{\substack{n=3\\m=2}}^{\infty}c_{nm}M_{nm},
	\end{equation}
	where the coefficients
	\begin{equation}
	c_{nm}=\frac{\left[ \mu_0^n\mu_1\mu_2 \left(\mu_1-\mu_2\right) -\mu_1^n\mu_0\mu_2 \left(\mu_0-\mu_2\right) + \mu_2^n\mu_0\mu_1\left(\mu_0-\mu_1\right)\right]
		\left[\nu_0^m\left(\nu_1-\nu_2\right) - \nu_1^m\left(\nu_0-\nu_2\right)+\nu_2^m\left(\nu_0-\nu_1\right)\right]}{m! n! p_{nm|Z}},
	\end{equation}
	can be shown to be non-negative for all pairs $(n,m)$~\cite{FedAlv}. Then, an upper bound on $M_{02}$ is straightforwardly given by
	\begin{equation}
	M_{02}\leq M_{02}^{\text{U}} = \frac{\Omega_{02}^{\text{U}}}{c_{02}},
	\end{equation}
	where we have lower bounded all the terms $c_{nm}M_{nm}$ in Eq.~(\ref{eq:M02sum}), with the exception of $c_{02}M_{02}$, by zero. Similarly, an upper bound on $M_{04}$ is directly given by
	\begin{equation}
	M_{04}\leq M_{04}^{\text{U}} = \frac{\Omega_{02}^{\text{U}}}{c_{04}}.
	\end{equation}

	\subsubsection{Upper bounds on  \texorpdfstring{$M_{20}$}{M20} and  \texorpdfstring{$M_{40}$}{M40}}
	By substituting $\kappa^{\mu}_A=1$, $\kappa^{\mu}_B=\mu$, $w^{\mu\nu}_A=(\mu-\nu)$, $w^{\nu_0\nu_1}_B=(\nu_0-\nu_1)\nu_0$ and $w^{\nu_1\nu_2}_B=(\nu_1-\nu_2)\nu_2$ in $\Omega$, we obtain the following function $\Omega_{20}$:
	\begin{equation}\label{eq:M20sum}
	\Omega_{20}:= \Omega =\sum_{n=2}^{\infty}c_{n0}M_{n0}+\sum_{\substack{n=2\\m=3}}^{\infty}c_{nm}M_{nm},
	\end{equation}
	where the coefficients
	\begin{equation}
	c_{nm}=\frac{\left[\mu_0^n\left(\mu_1-\mu_2\right) - \mu_1^n\left(\mu_0-\mu_2\right)+\mu_2^n\left(\mu_0-\mu_1\right)\right]
	    \left[ \nu_0^m\nu_1\nu_2 \left(\nu_1-\nu_2\right) -\nu_1^m\nu_0\nu_2 \left(\nu_0-\nu_2\right) + \nu_2^m\nu_0\nu_1\left(\nu_0-\nu_1\right)\right]
		}{m! n! p_{nm|Z}},
	\end{equation}
	can be shown to be non-negative for all the pairs $(n,m)$~\cite{FedAlv}. Then, an upper bound on $M_{20}$ is straightforwardly given by
	\begin{equation}
	M_{20}\leq M_{20}^{\text{U}} = \frac{\Omega_{20}^{\text{U}}}{c_{20}}.
	\end{equation}
	where we have lower bounded all the terms $c_{nm}M_{nm}$ in Eq.~(\ref{eq:M20sum}), with the exception of $c_{20}M_{20}$, by zero. Similarly, an upper bound on $M_{40}$ is directly given by
	\begin{equation}
	M_{40}\leq M_{40}^{\text{U}} = \frac{\Omega_{20}^{\text{U}}}{c_{40}}.
	\end{equation}

	\subsubsection{Upper bound on \texorpdfstring{$M_{13}$}{M13}}
	By substituting $\kappa^{\mu}_A=\kappa^{\mu}_B=1$, $w^{\mu\nu}_A=(\mu^2-\nu^2)$ and $w^{\mu\nu}_B=(\mu-\nu)$ in $\Omega$, we obtain a function $\Omega_{13}$ such that:
	\begin{equation}
	-\Omega_{13}:= \Omega =\sum_{m=2}^{\infty}c_{1m}M_{1m}+\sum_{\substack{n=3\\m=2}}^{\infty}c_{nm}M_{nm},
	\end{equation}
	where the coefficients
	\begin{equation}
	c_{nm}=-\frac{\left[\mu_0^n\left(\mu_1^2-\mu_2^2\right) -\mu_1^n \left(\mu_0^2-\mu_2^2\right)+\mu_2^n\left(\mu_0^2-\mu_1^2\right)\right] \left[\nu_0^m\left(\nu_1-\nu_2\right) - \nu_1^m\left(\nu_0-\nu_2\right)+\nu_2^m\left(\nu_0-\nu_1\right)\right]}{m! n! p_{nm|Z}},
	\end{equation}
	can be shown to be positive for the pairs $(n,m)\in S_+$, being $S_+=\{(n,m)|n=1,m\geq2\}$ and negative for the pairs $(n,m)\in S_-$, being $S_-=\{(n,m)|n\geq3,m\geq2\}$~\cite{FedAlv}. Then, by following a similar procedure to that used to derive Eq.~(\ref{eq:M11}), an upper bound on $M_{13}$ can be obtained as
	\begin{equation}
	M_{13}\leq M_{13}^{\text{U}} = \frac{c_{\max}(M_{z}-M_{0A}^L-M_{0B}^L+M_{00}^{\text{U}})-\Omega_{13}^L}{c_{13}+c_{\max}},
	\end{equation}
	{where $c_{\max}=\frac{\left(\nu_1-\nu_2\right)}{e^{-\nu_0}p_{\nu_0}}\frac{\left(\mu_1^2-\mu_2^2\right)}{e^{-\mu_0}p_{\mu_0}}$,} and $M_{0A}^L$ and $M_{0B}^L$ are given by Eqs.~(\ref{eq:M0AL}) and~(\ref{eq:M0BL}), respectively.

	\subsubsection{Upper bound on  \texorpdfstring{$M_{31}$}{M31}}
	By substituting $\kappa^{\mu}_A=\kappa^{\mu}_B=1$, $w^{\mu\nu}_A=(\mu-\nu)$ and $w^{\mu\nu}_B=(\mu^2-\nu^2)$ in $\Omega$, we obtain a function $\Omega_{31}$ such that:
	\begin{equation}
	-\Omega_{31}:= \Omega =\sum_{n=2}^{\infty}c_{n1}M_{n1}+\sum_{\substack{n=2\\m=3}}^{\infty}c_{nm}M_{nm},
	\end{equation}
	where the coefficients
	\begin{equation}
	c_{nm}=-\frac{ \left[\mu_0^n\left(\mu_1-\mu_2\right) - \mu_1^n\left(\mu_0-\mu_2\right)+\mu_2^n\left(\mu_0-\mu_1\right)\right]
    \left[\nu_0^m\left(\nu_1^2-\nu_2^2\right) -\nu_1^m \left(\nu_0^2-\nu_2^2\right)+\nu_2^m\left(\nu_0^2-\nu_1^2\right)\right]}{m! n! p_{nm|Z}},
	\end{equation}
	can be shown to be positive for the pairs $(n,m)\in S_+$, with $S_+=\{(n,m)|n\geq2,m=1\}$ and negative for the pairs $(n,m)\in S_-$, with $S_-=\{(n,m)|n\geq2,m\geq3\}$~\cite{FedAlv}. Then, by following a similar procedure to that used to derive Eq.~(\ref{eq:M11}), an upper bound on $M_{31}$ can be obtained as
	\begin{equation}
	M_{31}\leq M_{31}^{\text{U}} = \frac{c_{\max}(M_{z}-M_{0A}^L-M_{0B}^L+M_{00}^{\text{U}})-\Omega_{31}^L}{c_{31}+c_{\max}},
	\end{equation}
	{where $c_{\max}=\frac{\left(\mu_1-\mu_2\right)}{e^{-\mu_0}p_{\mu_0}}\frac{\left(\nu_1^2-\nu_2^2\right)}{e^{-\nu_0}p_{\nu_0}}$,} and $M_{0A}^L$ and $M_{0B}^L$ are given by Eqs.~(\ref{eq:M0AL}) and~(\ref{eq:M0BL}), respectively.

	\section{Channel model}
	\label{app:channel-model}
	
	For our simulations, we use the channel model of \cite{curty2019simple}, which we summarize here. We model the overall loss between Alice (Bob) and Charlie by a beamsplitter of transmittance $\sqrt{\eta}$, which includes the channel transmissivity and the quantum efficiency of Charlie's detectors. We consider that the quantum channels connecting Alice and Bob with Charlie introduce both phase and polarisation misalignments. We model the phase mismatch between Alice and Bob's pulses by shifting Bob's signals by an angle $\phi = \delta_{\textrm{ph}} \pi$. We model polarisation misalignment as a unitary operation that transforms Alice's (Bob's) polarisation input mode $a^{\dagger}_{\textrm{in}}$ ($b^{\dagger}_{\textrm{in}}$) into the orthogonal polarisation output modes $a^{\dagger}_{\textrm{out}}$ and $a^{\dagger}_{\textrm{out} \bot}$ ($b^{\dagger}_{\textrm{out}}$ and $b^{\dagger}_{\textrm{out} \bot}$) as follows: $a^{\dagger}_{\textrm{in}} \to \cos(\theta_A) a^{\dagger}_{\textrm{out}} - \sin(\theta_A) a^{\dagger}_{\textrm{out} \bot}$ ($b^{\dagger}_{\textrm{in}} \to \cos(\theta_B) b^{\dagger}_{\textrm{out}} - \sin(\theta_B) b^{\dagger}_{\textrm{out} \bot}$). The rotation angles are assumed to be $\theta_A = -\theta_B = \arcsin(\sqrt{\delta_{\textrm{pol}}})$.

	With this channel model, it can be shown \cite{curty2019simple} that the probability that Charlie reports a successful detection, given that both users employ the $X$ basis, is given by
	\begin{equation}
	Q_X = (1-p_d)(e^{-\gamma \Omega(\phi,\theta)} + e^{\gamma \Omega(\phi,\theta)}) e^{-\gamma} - 2 (1-p_d)^2 e^{-2 \gamma},
	\end{equation}
	where $\gamma = \sqrt{\eta} \alpha^2$, $\theta = \theta_A - \theta_B$, and $\Omega(\phi,\theta) = \cos \phi \cos \theta$. The probability that Alice and Bob end up with different key bits is given by
	\begin{equation}
	\label{eq:biterrorprob}
	e_X = \frac{e^{-\gamma \Omega(\phi,\theta)} - (1-p_d) e^{-\gamma}}{e^{-\gamma \Omega(\phi,\theta)} + e^{\gamma \Omega(\phi,\theta)} - 2 (1-p_d) e^{-\gamma}},
	\end{equation}
	while the probability that Charlie reports a successful detection, given that both users employ the $Z$ basis and select the intensities $\mu$ and $\nu$, respectively, is 
	\begin{equation}
	Q^{\mu \nu} = 2 (1-p_d) \left[e^{-\frac{(\mu^2+\nu^2) \sqrt{\eta}}{2}} I_0 (\mu \nu \sqrt{\eta} \cos \theta) - (1-p_d) e^{-(\mu^2+\nu^2) \sqrt{\eta}} \right],
	\end{equation}
	where $I_0 (z) = \frac{1}{2 \pi i} \oint e^{(z/2)(t+1/t)} t^{-1} \textrm{d}t$ is the modified Bessel function of the first kind.
	
	In our simulations, we assume that the observed measurement counts equal their expected value, that is, we set $M_X = N p_X^2 Q_X$ and $M^{\mu \nu} = N p_Z^2 p_{\mu} p_{\nu} Q^{\mu \nu}$, where $M^{\mu \nu}$ denotes the number of successful rounds in which Alice and Bob select the $Z$ basis and the intensities $\mu$ and $\nu$, respectively. Also, we assume that the bit-error rate of the sifted-key equals the probability given by \cref{eq:biterrorprob}.

	\section{Proof of Equation (\ref{eq:pherrorprob})} \label{app:proof-conditional}
	
	Let us consider the evolution that the initial quantum state $\ket{\Phi} = \ket{\phi}^{\otimes N}$, where $\ket{\phi} = \ket{\psi}_{A_cAa} \otimes \ket{\psi}_{B_cBb}$ and $\ket{\psi}$ is given by \cref{eq:vir_psi}, experiences before step (\ref{step:ancilla-measurement-same}) in Box 3, by taking into account all operations applied to it. After Eve's measurement in step 2 of Box 3, it is transformed to $\hat{M}_{\textrm{eve}} \ket{\Phi}$, where $\hat{M}_{\textrm{eve}}$ is the operator associated with her outcome. Let us reorder $\ket{\Phi}$ as $\ket{\Phi} = \ket{\phi}^{\otimes M} \ket{\phi}^{\otimes \bar{M}}$, writing first (last) the $M$ ($\bar{M}$) successful (unsuccessful) rounds. In the virtual sifting step, Alice and Bob measure all subsystems $A_c$ and $B_c$, using measurement operators $\{\hat{O}_s = \ketbra{00}{00} + \ketbra{11}{11}$, $\hat{O}_d = \mathds{I} - \hat{O}_s\}$. Again, let us reorder $\ket{\Phi} = \ket{\phi}^{\otimes M_s} \ket{\phi}^{\otimes M_d} \ket{\phi}^{\otimes \bar{M}_s} \ket{\phi}^{\otimes \bar{M}_d}$, writing first (second) the $M_s$ ($M_d$) successful rounds in which the users used the same (a different) basis, and third (fourth) the $\bar{M}_s$ ($\bar{M}_d$) unsuccessful rounds in which the users used the same (a different) basis. The unnormalised quantum state just before step \ref{step:ancilla-measurement-same} is then given by
	\begin{equation}
	\begin{aligned}
	\hat{O}_s^{\otimes M_s} \hat{O}_d^{\otimes M_d} \hat{O}_s^{\otimes \bar{M}_s} \hat{O}_d^{\otimes \bar{M}_d} \hat{M}_{\textrm{eve}} \ket{\Phi} &= \hat{M}_{\textrm{eve}} \hat{O}_s^{\otimes M_s} \hat{O}_d^{\otimes M_d} \hat{O}_s^{\otimes \bar{M}_s} \hat{O}_d^{\otimes \bar{M}_d} \ket{\Phi} \\ 
	&= \hat{M}_{\textrm{eve}} (\hat{O}_s \ket{\phi})^{\otimes M_s} (\hat{O}_d \ket{\phi})^{\otimes M_d}  (\hat{O}_s \ket{\phi})^{\otimes \bar{M}_s} (\hat{O}_d \ket{\phi})^{\otimes \bar{M}_d},
	\end{aligned}
	\end{equation}
	where we have used the fact that $\hat{M}_{\textrm{eve}}$ commutes with the sifting operators, as they act on different systems. Next, in step \ref{step:ancilla-measurement-same}, Alice and Bob measure the registers $A_c$, $B_c$, $A$ and $B$ for all rounds in $\mathcal{M}_s$, one by one. Let $u \in \{1,...,M_s\}$ index the rounds in $\mathcal{M}_s$, let $\xi_u$ be the outcome of the measurement of the $u$-th registers, and let $\hat{M}_u$ denote its associated measurement operator. Note that $\hat{M}_u \hat{O}_s = \hat{M}_u$. The unnormalised state just before their measurement of the $u$-th registers is
	\begin{equation}
	\ket{\Phi_u} = \hat{M}_{\textrm{eve}} (\otimes_{l=1}^{u-1} \hat{M}_l \ket{\phi}) (\hat{O}_s \ket{\phi}_u) (\hat{O}_s \ket{\phi})^{\otimes (M_s-u)} (\hat{O}_d \ket{\phi})^{\otimes M_d} (\hat{O}_s \ket{\phi})^{\otimes \bar{M}_s} (\hat{O}_d \ket{\phi})^{\otimes \bar{M}_d},
	\end{equation}
	where we have highlighted the initial quantum state of the $u$-th round, renaming it as $\ket{\phi}_u$. Since we are interested in the reduced state of the round $u$, we trace out the other rounds, which we denote by $\bar{u}$:
	\begin{equation}
	\hat{\sigma}_u = \Tr_{\bar{u}} [\ketbra{\Phi_u}] = \sum_{\vec{\bar{u}}} \braket{\vec{\bar{u}}}{\Phi_u}\! \braket{\Phi_u}{\vec{\bar{u}}} = \sum_{\vec{\bar{u}}} \hat{M}_{\vec{\bar{u}}} \hat{O}_s \ketbra{\phi}_u \hat{O}_s^{\dagger} \hat{M}_{\vec{\bar{u}}}^{\dagger} ,
	\end{equation}
	where
	\begin{equation}
	\hat{M}_{\vec{\bar{u}}} = \bra{\vec{\bar{u}}}\hat{M}_{\textrm{eve}} (\otimes_{l=1}^{u-1} \hat{M}_l \ket{\phi}) (\hat{O}_s \ket{\phi})^{\otimes (M_s-u)} (\hat{O}_d \ket{\phi})^{\otimes M_d} (\hat{O}_s \ket{\phi})^{\otimes \bar{M}_s} (\hat{O}_d \ket{\phi})^{\otimes \bar{M}_d},
	\end{equation}
	and the states $\{\ket{\vec{\bar{u}}}\}$ represent a basis for all the subsystems $A_c$, $A$, $B_c$, $B$, $a$ and $b$ of all the rounds in the protocol except the $u$-th round in $\mathcal{M}_s$ . The operator $\hat{\sigma}_u$ is unnormalised, and its trace denotes the joint probability of all previous measurement outcomes in the protocol. This includes Eve's measurement outcomes and Alice and Bob virtual sifting results, which we collectively denote as the event $\bm{\xi}$; as well as Alice and Bob's measurement outcomes $\xi_1,...,\xi_{u-1}$ of the previous $u-1$ registers. That is, $\Tr[\hat{\sigma}_u] = \Pr(\bm{\xi}, \xi_1,...,\xi_{u-1})$. The probability that Alice and Bob learn that they used the $Z$ basis and sent Fock states $\ket{n,m}$ in the $u$-th round of $\mathcal{M}_s$, conditioned on all the previous events, is
	\begin{equation}
	\label{eq:zz_na_nb}
	\begin{aligned}
	\Pr\left(\xi_u = Z_{n m}|\bm{\xi}, \xi_{1}, ..., \xi_{u-1}\right) &= \frac{\Tr \left[\bra{11}_{A_cB_c}\bra{n m}_{AB} \hat{\sigma}_u \ket{11}_{A_cB_c}\ket{n m}_{AB}\right]}{\Tr[\hat{\sigma}_u]} \\
	&= \frac{\Tr \left[\bra{11}_{A_cB_c}\bra{n m}_{AB} \sum_{\vec{\bar{u}}}  \hat{M}_{\vec{\bar{u}}} \ketbra {\phi_u }\hat{M}_{\vec{\bar{u}}}^{\dagger} \ket{11}_{A_cB_c}\ket{n m}_{AB}\right]}{\Pr(\bm{\xi}, \xi_1,...,\xi_{u-1})} \\
	&= \frac{\sum_{\vec{\bar{u}}} \norm{\hat{M}_{\vec{\bar{u}}} \bra{11}_{A_cB_c}\bra{n m}_{AB}   \ket{\phi}_u}^2}{\Pr(\bm{\xi}, \xi_1,...,\xi_{u-1})} \\
	&= \frac{p_Z^2 p_{nm|Z} \sum_{\vec{\bar{u}}} \norm{\hat{M}_{\vec{\bar{u}}} \ket{n}_a \ket{m}_b}^2 }{\Pr(\bm{\xi}, \xi_1,...,\xi_{u-1})}\\
	&= \frac{p_Z^2 p_{nm|Z}  \bra{n}_a \bra{m}_b \left(\sum_{\vec{\bar{u}}} \hat{M}_{\vec{\bar{u}}}^\dagger  \hat{M}_{\vec{\bar{u}}}\right)\ket{n}_a \ket{m}_b}{\Pr(\bm{\xi}, \xi_1,...,\xi_{u-1})},
	\end{aligned}
	\end{equation}
	where in the second equality we have used $\hat{O}_s \hat{M}_{\vec{\bar{u}}} = \hat{M}_{\vec{\bar{u}}} \hat{O}_s$ and $\hat{O}_s \ket{11}_{A_c B_c} = \ket{11}_{A_c B_c}$. Now, let $\hat{E}_{u} =\sum_{\vec{\bar{u}}} \hat{M}_{\vec{\bar{u}}}^\dagger  \hat{M}_{\vec{\bar{u}}}$. Since  $\hat{E}_{u}$ is a sum of positive semi-definite operators, it is positive semi-definite. Therefore, we can decompose it as $\hat{E}_{u} = \sqrt{\hat{E}_{u}} \sqrt{\hat{E}_{u}}$, and rewrite \cref{eq:zz_na_nb} as
	\begin{equation}
	\label{eq:zz_na_nb2}
	\begin{aligned}
	\Pr\left(\xi_u = Z_{n m}|\bm{\xi}, \xi_{1}, ..., \xi_{u-1}\right) &=  \frac{p_Z^2 p_{nm|Z}  \bra{n}_a \bra{m}_b \sqrt{\hat{E}_{u}} \sqrt{\hat{E}_{u}} \ket{n}_a \ket{m}_b}{\Pr(\bm{\xi}, \xi_1,...,\xi_{u-1})} \\
	&= \frac{p_Z^2 p_{nm|Z} \norm{\sqrt{\hat{E}_{u}} \ket{n}_a \ket{m}_b}^2 }{\Pr(\bm{\xi}, \xi_1,...,\xi_{u-1})}.
	\end{aligned}
	\end{equation}
	
	Using an identical approach, we can show that the probability that Alice and Bob will learn that they used the $X$ basis and sent cat states $\ket{C_i}\ket{C_j}$ in the $u$-th successful round is
	\begin{equation}
	\label{eq:xx_ij}
	\Pr\left(\xi_u = X_{i j}|\bm{\xi}, \xi_{1}, ..., \xi_{u-1}\right) = \frac{p_X^2  \norm{\sqrt{\hat{E}_{u}} \ket{C_i}_a \ket{C_j}_b}^2 }{\Pr(\bm{\xi}, \xi_1,...,\xi_{u-1})}.
	\end{equation}

	Now, we want to relate the probability terms on the left hand side of \cref{eq:zz_na_nb2,eq:xx_ij}. For this, we use the approach of \cite{curty2019simple} and apply the Cauchy-Schwartz inequality to show that
	\begin{equation}
	\begin{aligned}
	\norm{\sqrt{\hat{E}_{u}} \ket{C_i}_a \ket{C_j}_b}^2 & = \sum_{\substack{n,n' \in \mathds{N}_i \\ m,m' \in \mathds{N}_j}} \sqrt{p_{n'm'|X}} \sqrt{p_{nm|X}} \bra{n'}_a \bra{m'}_b \sqrt{\hat{E}_{u}} \sqrt{\hat{E}_{u}} \ket{n} \ket{m} \\
	&\leq   \sum_{\substack{n,n' \in \mathds{N}_i \\ m,m' \in \mathds{N}_j}}  \sqrt{p_{n'm'|X}} \sqrt{p_{nm|X}} \norm{\sqrt{\hat{E}_{u}} \ket{n'}_a \ket{m'}_b}  \norm{\sqrt{\hat{E}_{u}}  \ket{n}_a \ket{m}_b} \\
	&= \left[ \sum_{n \in \mathds{N}_i, m \in \mathds{N}_j} \sqrt{p_{nm|X}}   \norm{\sqrt{\hat{E}_{u}}  \ket{n}_a \ket{m}_b} \right]^2.
	\end{aligned}	
	\end{equation}
	
	Combining the three previous equations, we obtain
	\begin{equation} \label{eq:PcondX-condZ}
	\begin{aligned}
	P \left(\xi_u = X_{i j} |\bm{\xi},\xi_1, ..., \xi_{u-1}\right) &\leq \frac{p_X^2  \left[\sum_{n\in \mathds{N}_i, m \in \mathds{N}_j} \sqrt{p_{nm|X}} \norm{\sqrt{\hat{E}_{u}}  \ket{n}_a \ket{m}_b} \right]^2  }{\Pr(\bm{\xi},\xi_1, ..., \xi_{u-1})} \\
	&=  \frac{p_X^2}{p_Z^2} \Bigg[~ \sum_{n\in \mathds{N}_i, m \in \mathds{N}_j} \sqrt{\frac{p_{nm|X}}{p_{nm|Z}} \Pr \left(\xi_u = Z_{n m}|\bm{\xi},\xi_1, ..., \xi_{u-1}\right)} ~\Bigg]^2,
	\end{aligned}
	\end{equation}
	and since a phase error occurs when $\xi_u \in \{X_{00}, X_{11}\}$, its probability is upper-bounded by
	\begin{equation} \label{eq:pherrorprob-app}
	\Pr \left(\xi_u \in \{X_{00}, X_{11}\}|\bm{\xi}, \xi_{0}, ..., \xi_{u-1}\right) \leq \frac{p_X^2}{p_Z^2} \sum_{j=0}^{1} \Bigg[~ \sum_{n, m \in \mathds{N}_j} \sqrt{\frac{p_{nm|X}}{p_{nm|Z}} \Pr \left(\xi_u = Z_{n m}|\bm{\xi},\xi_1, ..., \xi_{u-1}\right)} ~\Bigg]^2.
	\end{equation}
	Note that, since all probabilities are conditioned on $\bm{\xi}$, we can remove it from the conditions and work on the probability space in which the event $\bm{\xi}$ has happened. Also, to match the notation in \cref{eq:azumaNph,eq:azumaMnm,eq:azumaM00}, we rewrite \cref{eq:pherrorprob-app} as
	\begin{equation}
	\Pr \left(\xi_u \in \{X_{00}, X_{11}\}|\mathcal{F}_{u-1}\right) \leq \frac{p_X^2}{p_Z^2} \sum_{j=0}^{1} \Bigg[~ \sum_{n, m \in \mathds{N}_j} \sqrt{\frac{p_{nm|X}}{p_{nm|Z}} \Pr \left(\xi_u = Z_{n m}|\mathcal{F}_{u-1}\right)} ~\Bigg]^2,
	\end{equation}
	where $\mathcal{F}_{u-1}$ is the $\sigma$-algebra generated by $\xi_1,...,\xi_{u-1}$.

	\section{Inverse multiplicative Chernoff bound}
	\label{app:chernoff}
	
	Here, we state the result that we use to obtain the lower and upper bounds required in \cref{eq:linearprogram}. Let $\chi = \sum_{i=1}^{n} \chi_i$ be the outcome of a sum of $n$ independent Bernoulli random variables $\chi_i \in \{0,1\}$. Given the observation of the outcome $\chi$, its expectation value $\mathbb{E}[\chi]$ can be lower and upper bounded by \cite{zhang2017improved}
	\begin{equation}
	\label{eq:chernoff-begin}
	\begin{gathered}
	\mathbb{E}^L[\chi] = \frac{\chi}{1+\delta^L}, \\
	\mathbb{E}^{\text{U}}[\chi] = \frac{\chi}{1-\delta^{\text{U}}}, 
	\end{gathered}
	\end{equation}
	except with a probability $\varepsilon_c$, where $\delta^L$ and $\delta^{\text{U}}$ are the solutions of the following equations
	\begin{equation}
	\label{eq:chernoff-eqs}
	\begin{gathered}
	\left[\frac{e^{\delta^L}}{(1+\delta^L)^{1+\delta^L}}\right]^{\chi/(1+\delta^L)} = \frac{1}{2} \varepsilon_c \\
	\left[\frac{e^{-\delta^{\text{U}}}}{(1-\delta^{\text{U}})^{1-\delta^{\text{U}}}}\right]^{\chi/(1-\delta^{\text{U}})} = \frac{1}{2} \varepsilon_c.
	\end{gathered}
	\end{equation}
	The solutions to \cref{eq:chernoff-eqs} satisfy \cite{bahrani2019wavelength}
	\begin{equation}
	\label{eq:lambert}
	\begin{gathered}
	\frac{1}{1+\delta_L} =  W_0(-e^{\ln(\varepsilon_c/2-\chi)/\chi}), \\
	\frac{1}{1-\delta_U} = W_{-1}(-e^{\ln(\varepsilon_c/2-\chi)/\chi}),
	\end{gathered}
	\end{equation}
	where $W_0$ and $W_{-1}$ are branches of the Lambert $W$ function, which is the inverse of the function $f(z) = z e^z$.

		\clearpage
	\bibliography{refs}

\end{document}